\documentstyle[aps,eqsecnum,preprint,prd,psfig]{revtex}

\tighten
\begin{document}
\draft
\preprint{MPI--Ph/93--51/R; TUM--T31--42/93/R}
\title{
NEXT--TO--LEADING ORDER CORRECTIONS TO MESON MASSES IN THE HEAVY QUARK
EFFECTIVE THEORY
}
\author{Patricia Ball}
\address{
Physik--Department, TU M\"unchen, D--85747 Garching, FRG
}
\author{V.M.~Braun\thanks{On leave of absence from St.\ Petersburg
Nuclear Physics Institute, 188350 Gatchina, Russia.}}
\address{
Max--Planck--Institut f\"ur Physik, P.O.\ Box 40 12 12, D--80805
M\"unchen, FRG
}
\bigskip
\date{September 3, 1993}
\maketitle
\begin{abstract}
We use the QCD sum rule approach to calculate the splitting
between vector and pseudoscalar mesons containing one light and
one heavy quark, and the kinetic energy of the heavy quark.
Our result for the splitting induced by the chromomagnetic
interaction agrees to the experimental data on charm and beauty
mesons. For the matrix element of the kinetic energy operator,
we obtain the value $K=-(0.60\pm 0.10)\, {\rm GeV}^2$.
\end{abstract}
\section{Introduction}
In recent years there has been a continuous interest in the study of
mesons built of one heavy and one light quark. This interest is
fuelled by a constant flow of new experimental data and the results
of lattice calculations. The main theoretical achievement in the past
few years was the development of the heavy quark effective theory
(HQET, see \cite{HQET} for reviews), which provides one with a
systematic tool for the
study of the heavy quark limit $m_Q\rightarrow \infty$ and for the
classification of corrections which are suppressed by powers of the
heavy quark mass. Physical applications of this approach include
various decays of charm and beauty mesons, for which the
preasymptotical corrections $\sim 1/m_Q$ are likely to be
significant. Considerable effort has been made in the past two years
to estimate them, see e.g.\ \cite{latt,BBBD,N92,ball}, but this task
is far from being completed yet.
The Lagrangian density, written in terms of the
effective heavy quark fields, contains to  $O(1/m_Q)$ accuracy two
additional contributions \cite{FGL91,MRR92} apart from the leading
one. They are related in an obvious way to the non--relativistic
kinetic energy operator, and to the Pauli term, describing the
chromomagnetic interaction.
Matrix elements of these two operators over meson states
are fundamental observables in the heavy quark effective theory,
and determine the next--to--leading order corrections to the meson
masses suppressed by powers of the quark mass. The matrix element of
the chromomagnetic interaction operator is a leading contribution to
the mass splitting between vector and pseudoscalar mesons, and can
be directly related to the data on the meson spectrum.
The matrix element of the kinetic energy operator contains
information about the smearing in heavy quark momentum, which
is important in many applications and contributes, e.g., to
subleading form factors of semileptonic decays \cite{falk}, and
recently was shown to give a significant contribution to the electron
spectrum in inclusive B--decays \cite{BUV92}.

Up to now, there
exist no quantitative estimates for the kinetic energy of the
heavy quark, except for a single attempt in \cite{N92}.
This task proves to be difficult for lattice
QCD, because of  power divergences which need to be subtracted,
see e.g. \cite{MMS92}. In this paper we calculate the kinetic energy
and the chromomagnetic mass splitting using the QCD sum rule approach
\cite{SVZ}, including radiative corrections.

The presentation is organized as follows. \mbox{Sec.\ II} is
introductory, and contains a short discussion of the heavy quark
effective theory, giving necessary definitions and notations.
Some results of more technical nature are given in
\mbox{App.\ \ref{app:C}}. The QCD sum rules for the relevant
three--point functions in the effective theory are derived in
\mbox{Sec.\ III}. \mbox{Sec.\ IV} contains our main result, namely
the sum rule for the matrix element of the kinetic energy operator,
including two--loop radiative corrections and renormalization group
improvement to two--loop accuracy. Details of the calculation are
given in \mbox{Apps.\ \ref{app:A}} and \ref{app:B}.
Finally, in \mbox{Sec.\ V} we give the summary and conclusions.

\section{The Heavy Quark Expansion}

The dynamics of hadrons containing both light and heavy quarks
can be described by an effective field theory, in which the
heavy degrees of freedom are integrated out, and the resulting
effective Lagrangian is expanded in inverse powers of the heavy
quark mass. Following \cite{georgi} we introduce the
heavy quark effective field $h_v$ as
\begin{equation}
    P_+ Q(x) = e^{-im_Q(v\cdot x)}h_v(x),
\end{equation}
where
\begin{equation}
    P_+ = \frac{1}{2}(1+\not\! v)
\end{equation}
is the projector on the upper components of the heavy quark
field $Q(x)$, $m_Q$ is the heavy quark mass, and $v$ its
four--velocity.
The contributions of lower components of the Q--quark field
are integrated out, producing an effective theory with the
Lagrangian density \cite{FGL91,MRR92}
\begin{equation}
 {\cal L}  =  \bar h_v i(v\cdot D)h_v + \frac{1}{2m_Q}{\cal K}
 +\frac{1}{2m_Q}{\cal S} + O(1/m_Q^2),
\label{Lagrangian}
\end{equation}
where
\begin{equation}
 {\cal K}  =  \bar h_v (iD^\perp)^2 h_v
\label{kinetic}
 \end{equation}
is the operator of the non--relativistic kinetic energy, and
\begin{equation}
{\cal S}  =   \frac{1}{2}
 \left(\frac{\alpha_s(m_Q)}{\alpha_s(\mu)}\right)^{3/\beta_0}
\bar h_v \sigma_{\mu\nu}gF^{\mu\nu} h_v,
\label{magnetic}
 \end{equation}
 is the Pauli term.
 The covariant derivative is defined as
$ D_\mu = \partial_\mu -igA_\mu $.
In (\ref{kinetic}) we have introduced projectors on the directions
orthogonal to the velocity $v_\mu$:
\begin{eqnarray}
     g_{\mu\nu}^\perp &=& g_{\mu\nu}-v_\mu v_\nu,
  \nonumber\\
    (D^\perp)^2 &=& D_\mu D^\mu - (vD)^2.
\end{eqnarray}

The operator of  the chromomagnetic interaction
has  a non--trivial anomalous dimension, and the coefficient in front
of it receives corrections of higher orders in the coupling constant,
$C_{P}(\mu=m_Q) = 1/2 + O(\alpha_s(m_Q))$.
In contrast to this, the kinetic energy operator has zero anomalous
dimension to all orders, and its coefficient is exactly
one, cf.\ \cite{LM92}. This property follows from a residual
symmetry of the effective theory under non--relativistic boosts,
and is related to the fact that
the kinetic energy term in the non--relativistic Hamiltonian is
invariant under these transformations.

Operators in the full theory (QCD) are expanded in a series of
operators in the effective theory as
\begin{equation}
   O^{\text{QCD}} =\sum_i C_i(m_Q/\mu, \alpha_s(\mu))
O^{\text{HQET}}_i(\mu).
\end{equation}
The coefficients in this expansion depend on the scale $\mu$
according to the renormalization group equations,
\begin{equation}
 C(m_Q/\mu, \alpha_s(\mu))=   C(1,\alpha_s(m_Q))
\exp\left[ -\int_{g(\mu)}^{g(m_Q)}\frac{\gamma(g)}{\beta(g)} d g
    \right]   \,,
\end{equation}
 where $\gamma(g)$ is the anomalous dimension of the corresponding
 operator in the effective theory,
\begin{equation}
  \gamma = \gamma_0\, \frac{g^2}{16\pi^2} +\gamma_1
  \left(\frac{g^2}{16\pi^2}\right)^2 + \ldots \,,
\end{equation}
and $\beta(g)$ is the Gell-Mann--Low function
\begin{equation}
  \beta = -g\left[
  \beta_0\, \frac{g^2}{16\pi^2} +\beta_1
  \left(\frac{g^2}{16\pi^2}\right)^2 + \ldots \right] \,.
\end{equation}
At the scale $\mu=m_Q$, the coefficient
functions $C_i$ are determined
{}from the condition that the matrix elements of effective operators,
times the appropriate coefficient functions, equal
the matrix elements of the corresponding QCD operators at this scale
to the required accuracy.
For the practically important cases of  vector and axial currents
built of one heavy and one light quark, the heavy quark expansion
reads \cite{N92,JM}
\begin{eqnarray}
  \bar q \gamma_\mu Q &=& \left( 1-C_F \frac{\alpha_s(m_Q)}{\pi}
                      \right) (\bar q \gamma_\mu h_v)^{(\mu=m_Q)}
       +\frac{1}{2} C_F  \frac{\alpha_s(m_Q)}{\pi}
       (\bar q v_\mu h_v)^{(\mu=m_Q)}
\nonumber\\
  &&\mbox{}+\frac{1}{2m_Q} (\bar q \gamma_\mu i \not\!\!D^\perp
  h_v)^{(\mu=m_Q)} +O\left(\frac{1}{2m_Q} \frac{\alpha_s(m_Q)}{\pi}
     \right)\,,
\nonumber\\
  \bar q \gamma_\mu \gamma_5
  Q &=& \left( 1-C_F \frac{\alpha_s(m_Q)}{\pi}
        \right) (\bar q \gamma_\mu \gamma_5 h_v)^{(\mu=m_Q)}
       -\frac{1}{2} C_F  \frac{\alpha_s(m_Q)}{\pi}
       (\bar q v_\mu \gamma_5 h_v)^{(\mu=m_Q)}
\nonumber\\
  &&\mbox{}+\frac{1}{2m_Q} (\bar q \gamma_\mu i \not\!\!D^\perp
  h_v)^{(\mu=m_Q)} +O\left(\frac{1}{2m_Q} \frac{\alpha_s(m_Q)}{\pi}
     \right)\,,
\label{matching}
\end{eqnarray}
where $C_F = (N_c^2-1)/(2N_c)$.
Invariant functions and matrix elements of operators in HQET are
most conveniently defined using the so--called trace formalism
\cite{traceformalism}, which makes the spin symmetries explicit.
Following \cite{N92}, we define
\begin{equation}
\langle 0 | \bar q \Gamma h_v |M(v)\rangle =
\frac{1}{2} F(\mu) \mbox{Tr} \{ \Gamma {\cal M}(v)\},
\label{fhql}
\end{equation}
where ${\cal M}(v)$ is the spin wave function
\begin{equation}
{\cal M}(v) = \sqrt{m_M} P_+
\left \{ \begin{array} {cc}
 -i \gamma_5 & {\rm for\ }J^P = 0^-,\\
 \not\!\epsilon &  {\rm for\ }J^P = 1^-.
\end{array} \right.
\end{equation}
The relation in (\ref{fhql}) is valid for an arbitrary structure of
Dirac matrices $\Gamma$.
To two--loop accuracy we have
\begin{equation}
(\bar q \Gamma h_v)^{(\mu=m_Q)} =(\bar q \Gamma h_v)^{(\mu)}
\left(\frac{\alpha_s(m_Q)}{\alpha_s(\mu)}\right)^{\gamma_0/(2\beta_0)}
\left[ 1 +\frac{\gamma_0}{8\beta_0}\left(
\frac{\gamma_1}{\gamma_0}-\frac{\beta_1}{\beta_0} \right)
\frac{\alpha_s(m_Q)-\alpha_s(\mu)}{\pi}\right]
\label{running}
\end{equation}
with the anomalous dimensions \cite{JM,VS87,BG91}
\begin{eqnarray}
\gamma_0 = -4\,, &\quad &\gamma_1 =-254/9-56\pi^2/27 +20 n_f/9\,,
\nonumber\\
\beta_0 = 11-2n_f/3\,, &\quad & \beta_1 = 102-38n_f/3 \,.
\label{anomdim}
\end{eqnarray}
The effective coupling $F(\mu)$ depends on the scale in a similar
way.
Defining physical lepton decay constants, e.g. for B--mesons,
by
 \begin{eqnarray}
 \langle 0| \bar q \gamma_\mu \gamma_5 b|B(q)\rangle
   &=& i f_B q_\mu,
\nonumber\\
  \langle 0| \bar q \gamma_\mu  b|B^*(q,\epsilon)\rangle
 &=& f_{B^*} m_{B^*} \epsilon_\mu, \hspace{1cm}
  \epsilon\cdot q=0, \epsilon^2=-1,
\label{couplings}
 \end{eqnarray}
and taking into account the matching conditions in (\ref{matching}),
one obtains the relations of the physical couplings to the effective
ones \cite{BBBD,JM}:
\begin{eqnarray}
f_B\sqrt{m_B} &=& \left(1-\frac{2}{3}\frac{\alpha_s(m_B)}{\pi}\right)
               F(\mu=m_B)+O(1/m_B),
\nonumber\\
f_{B^*}\sqrt{m_{B^*}} &=&
\left(1-\frac{4}{3}\frac{\alpha_s(m_B)}{\pi}\right)
               F(\mu=m_B) +O(1/m_B).
\end{eqnarray}
Furthermore, we define matrix elements of the operators of kinetic
energy and chromomagnetic interaction over heavy--light mesons as
\begin{eqnarray}
\langle M(v)|{\cal K}|M(v)\rangle &=& - K\,
\mbox{Tr}\{\bar{\cal M}(v){\cal M}(v)\}\,,
\nonumber\\
\langle M(v)|{\cal S}|M(v)\rangle &=& -d_M \Sigma\,
\mbox{Tr}\{\bar{\cal M}(v){\cal M}(v)\} \,.
\end{eqnarray}
Here $d_M$ is defined by
\begin{equation}
P_+ \sigma_{\alpha\beta}{\cal M}(v)\sigma^{\alpha\beta}
 =  2 d_M{\cal M}(v),
\end{equation}
yielding
 \begin{equation}
   d_M =
\left \{ \begin{array}{cc}
 3 & {\rm for\ }J^P = 0^-,\\
 -1 & {\rm for\ }J^P = 1^-.
\end{array} \right.
\end{equation}
Note that the normalization of matrix elements is
\begin{equation}
 \langle M(v)|\bar h_v h_v|M(v)\rangle = 2\cdot\left( -\frac{1}{2}
\right) \mbox{Tr}\{\bar{\cal M}(v){\cal M}(v)\} = 2 m_M \,,
\end{equation}
so that
\begin{eqnarray}
 K &=&\frac{\langle M(v)|{\cal K}|M(v)\rangle}{
\langle M(v)|\bar h_v h_v|M(v)\rangle},
\nonumber\\
 \Sigma &=& \frac{1}{d_M}\frac{\langle M(v)|{\cal S}|M(v)\rangle}{
 \langle M(v)|\bar h_v h_v|M(v)\rangle}.
\end{eqnarray}
To $1/m_Q$ accuracy the  meson masses are given by
\begin{equation}
  m_M  =  m_Q +\bar\Lambda -\frac{1}{2m_Q}[K +d_M \Sigma],
\label{mascor}
\end{equation}
where $\bar\Lambda$ determines the difference between the quark and
meson masses in the heavy quark limit, and is a fundamental
observable in the heavy quark effective theory.
It has received
a lot of attention recently, see e.g. \cite{BBBD,N92}.
The chromomagnetic interaction
determines the splitting between pseudoscalar and vector mesons:
\begin{equation}
m_V^2 -m_P^2 = 4\Sigma +O(1/m_Q).
\label{splitting}
\end{equation}
The relation (\ref{mascor}) can be derived in a variety of ways
and is widely known \cite{falk,BUV92}. One possibility is, e.g.,
to consider the $1/m_Q$ expansion of suitable two--point
correlation functions in QCD near the particle--type singularity,
corresponding to the lowest--lying meson state, cf.\
\mbox{App.\ \ref{app:C}}.

\section{The Sum Rules}

The aim of this paper is to obtain quantitative estimates of the matrix
elements $K$ and $\Sigma$. We use the method of QCD sum rules, see the
book \cite{shifman}, which we apply directly to suitable correlation
functions in HQET. In particular, we consider the following
three--point correlation functions at zero--recoil:
\begin{eqnarray}
\lefteqn{
i^2\int dx\int dy\, e^{i\omega (v\cdot x)-i\omega' (v\cdot y)}
 \langle 0|T\left\{\bar q(x)\Gamma_1 h_v(x)  {\cal K}(0) \bar h_v(y)
 \Gamma_2 q(y)\right\}|0\rangle          =}
\nonumber\\
&&\mbox{}=- \mbox{Tr}\{\Gamma_1 P_+ \Gamma_2 \}
T_K(\omega,\omega'),  \hspace{5cm}
\label{CFK}
\end{eqnarray}
\begin{eqnarray}
\lefteqn{
i^2\int dx\int dy\, e^{i\omega (v\cdot x)-i\omega' (v\cdot y)}
 \langle 0|T\left\{\bar q(x)\Gamma_1 h_v(x)  {\cal S}(0) \bar h_v(y)
 \Gamma_2 q(y)\right\}|0\rangle =  }
\nonumber\\
&&\mbox{}=  - d_M \mbox{Tr}\{\Gamma_1 P_+ \Gamma_2 \}
T_\Sigma(\omega,\omega'). \hspace{5cm}
\label{CFS}
\end{eqnarray}

Saturating the three--point functions with hadron states,
one can isolate the contribution of interest as the one having
poles in both the variables $\omega$ and $\omega'$ at the value
$\omega = \omega' =\bar\Lambda$:
\begin{eqnarray}
  T_\Sigma(\omega,\omega')&=&
\frac{\Sigma(\mu) F^2(\mu)}
{4(\bar\Lambda-\omega)(\bar\Lambda-\omega')}
+\ldots
\nonumber\\
  T_K(\omega,\omega')&=&
 \frac{K F^2(\mu)}{4(\bar\Lambda-\omega)(\bar\Lambda-\omega')}
+\ldots
\label{poles}
\end{eqnarray}
where the scale $\mu$ is the normalization point of the
currents, and
$F^2(\mu)$ is the coupling (squared) of the effective current
to the lowest--lying meson state, defined in (\ref{fhql}).
This coupling, in turn, can be
obtained as the residue of the pole at $\omega=\bar\Lambda$ in
the two--point correlation function
\begin{equation}
i\int d^4x\,e^{i\omega (v\cdot x)}\langle 0 |  T \left\{ \bar q(x)
 \Gamma_1 h_v(x) \bar h_v(0) \Gamma_2 q(0)
 \right\}|0\rangle
   = -\frac{1}{2} \mbox{Tr}\{\Gamma_1 P_+ \Gamma_2 \} \Pi(\omega),
\label{CF2-def}
\end{equation}
\begin{equation}
\Pi(\omega)  =
\frac{F^2(\mu)}{2(\bar\Lambda-\omega)} +\ldots
\label{CF2}
\end{equation}
To derive Eqs.\ (\ref{poles}) and (\ref{CF2}), the following relation
is useful \cite{N92}:
\begin{equation}
\sum_{\text{polarizations}} \mbox{Tr}\{\Gamma_1 {\cal M}(v)\}
 \mbox{Tr}\{\bar{\cal M}(v)\Gamma_2\}
 = -2 m_M  \mbox{Tr}\{\Gamma_1 P_+ \Gamma_2\}.
\label{polarize}
\end{equation}
For example, for the lowest--lying meson contribution to
the correlation function $\Pi(\omega)$, one obtains
\begin{equation}
\left(\frac{1}{2} F(\mu)\right)^2
\mbox{Tr}\{\Gamma_1 {\cal M}(v)\}\,
\frac{1}{2m_M(\bar\Lambda -\omega)}\,
\mbox{Tr}\{\bar{\cal M}(v)\Gamma_2\}
= \frac{F^2(\mu)}{2(\bar\Lambda -\omega)}
\left(-\frac{1}{2}\right) \mbox{Tr}\{\Gamma_1 P_+ \Gamma_2\},
\end{equation}
which is the result given in (\ref{CF2}).

\begin{figure}
$$
\ \psfig{file=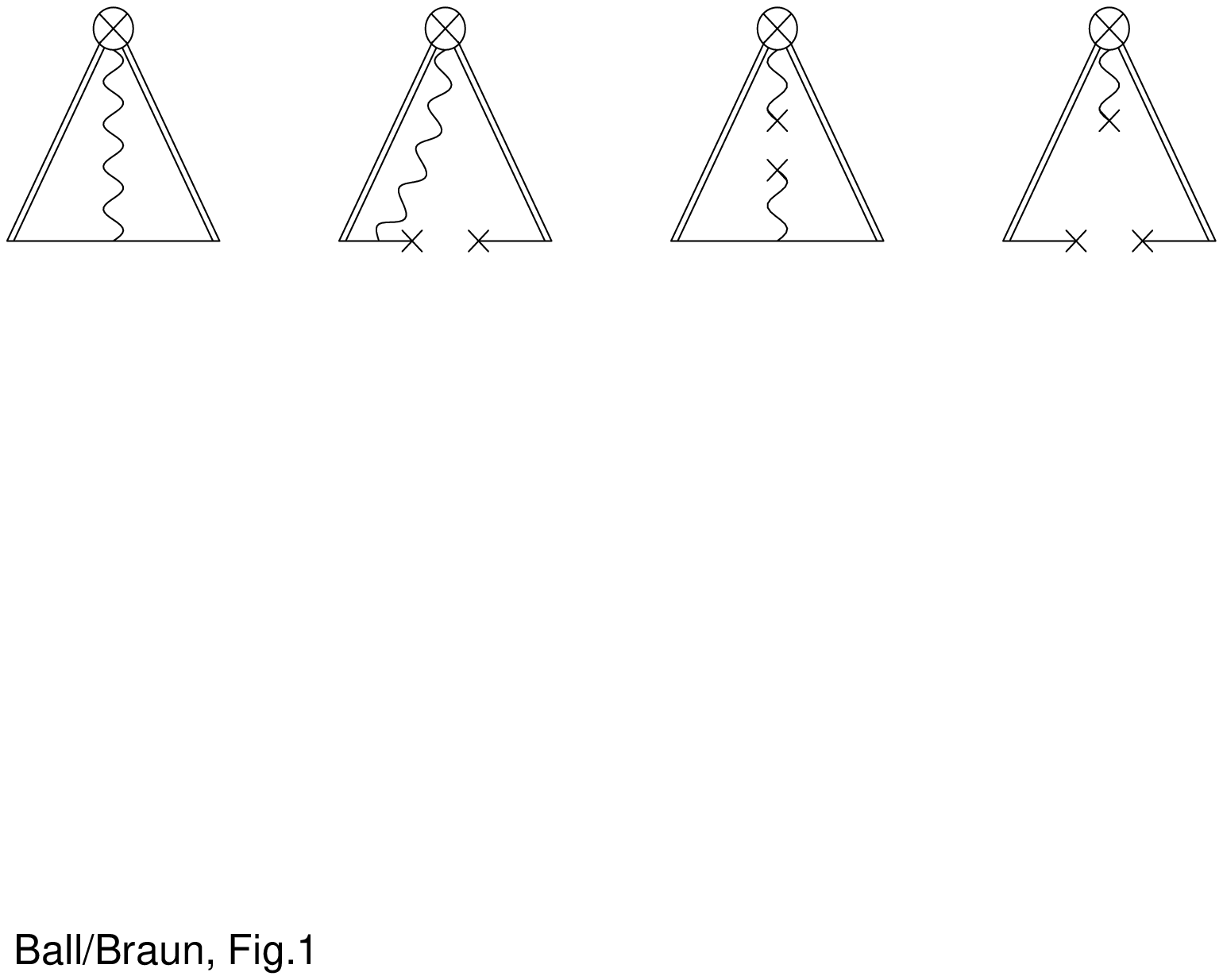,height=7in}
$$
\caption{The diagrams contributing to $\rho_\Sigma$,
Eq.~(\protect{\ref{spectr}}).\label{fig:1}}
\end{figure}

The correlation functions
$\Pi(\omega)$, $T_\Sigma(\omega,\omega')$, and $T_K(\omega,\omega')$
can be calculated in the Euclidian region, for negative
$\omega,\omega'$, and receive contributions from perturbation theory
and from vacuum condensates. The results can be written in form
of (double) dispersion relations,
\begin{eqnarray}
\Pi(\omega) &=& \int\!\! \frac{ds}{s-\omega}\, \rho_{\Pi}(s),
\nonumber\\
T_\Sigma(\omega,\omega') &=& \int\!\!\int\!\!\frac{ds}{s-\omega}\,
\frac{ds'}{s'-\omega'}\, \rho_{\Sigma}(s,s'),
\nonumber\\
 T_K(\omega,\omega') &=& \int\!\!\int\!\!\frac{ds}{s-\omega}\,
 \frac{ds'}{s'-\omega'}\, \rho_K(s,s'),
\label{disp}
\end{eqnarray}
where the spectral densities are subject to direct calculation
in HQET:
\begin{equation}\label{eq:decomposition}
\rho = \rho^{\text{pert}}\, \langle \openone \rangle + \rho^{(3)}\,
\langle \bar q q\rangle + \rho^{(4)}\,\left\langle
\frac{\alpha_s}{\pi} G^2\right\rangle + \rho^{(5)}\,\langle \bar q
\sigma g G q \rangle + \ldots
\end{equation}
The relevant Feynman diagrams, which contribute to the
three--point correlation functions to first order in $\alpha_s$,
are shown in \mbox{Fig.\ 1} for the chromomagnetic operator,
and in \mbox{Figs.\ 2, 3, and 4} for the kinetic energy (up to
dimension 4).
 The graphs missing there turn out
to be vanishing. Note that the leading contributions to the
correlation function $T_\Sigma$, Fig.~\ref{fig:1},
are of $O(\alpha_s)$. On the contrary, the leading perturbative
contribution to the kinetic energy is of order $O(1)$, see Fig.~2(a).

In this section, we consider the sum rule for the
chromomagnetic mass splitting, taking into account the set of
diagrams in Fig.~1, and the leading order sum rule for the kinetic
energy, taking into account the perturbative contribution in
Fig.~2(a) and the leading non--perturbative correction, which in this
case is due to the mixed quark--gluon condensate. The contribution
to the kinetic energy of the quark condensate is of $O(\alpha_s)$.
 The full sum rule for the kinetic energy,
with account for all the contributions in Figs.~2, 3, and 4 and with
renormalization group improvement to  two--loop accuracy, is
considered in detail in the next section.

A straightforward calculation yields:
\begin{eqnarray}
 \rho_{\Pi}(s) &=& \frac{3}{2\pi^2} s^2 -
 \frac{1}{2}\langle\bar q q\rangle\delta(s)
 +\frac{1}{32}\langle \bar q \sigma gG q\rangle \delta''(s) +
O(\alpha_s),\nonumber\\
\rho_K(s,s') &=&  -\frac{3}{4\pi^2} s^4 \delta(s-s') +\frac{3}{32}
\langle \bar q \sigma gG q\rangle \delta(s)\delta(s'),
\nonumber\\
\rho_{\Sigma}(s,s') &=&
\frac{\alpha_s}{\pi^3}
\left\{ s'^2(s-s'/3)\theta(s-s') + s^2(s'-s/3)\theta(s'-s)\right\}
\nonumber\\
&&\mbox{}-
\frac{\alpha_s}{3\pi}\langle\bar q q\rangle
\{s\delta(s')+s'\delta(s)\}
+\frac{1}{48}\left\langle\frac{\alpha_s}{\pi} G^2
\right\rangle\delta(s-s')
-\frac{1}{48}\langle \bar q \sigma gG q\rangle \delta(s)\delta(s').
\label{spectr}
\end{eqnarray}
Following the usual strategy of the QCD sum rule approach, we
apply a Borel transformation to the correlation functions in both the
external momenta $\omega$ and $\omega'$:
\begin{equation}
\hat B_\omega(t)\,
\frac{1}{\bar\Lambda-\omega}=
\frac{1}{t} \exp(-\bar\Lambda/t).\label{Borel}
\end{equation}
The transformation $\hat{B}_\omega(t)$ introduces the Borel parameter
$t$ instead of the external momenta.

We next equate the representations of the correlation functions
in terms of hadronic states, Eqs.~(\ref{poles}) and (\ref{CF2}), and
in terms of spectral densities, Eqs.~(\ref{disp}) and
(\ref{spectr}). In the latter we constrain the region of integration
to the interval of duality $0<s,s'<\omega_0$. Due to the symmetry of
the correlation functions, it is natural to take the two
Borel parameters in the three--point functions equal to each other,
$t_1=t_2$.  The normalization of the Isgur--Wise function
at zero recoil \cite{IW89}, further requires that
the value of the Borel parameter in the three--point function
takes twice the value as in the two--point function,
$t_1=t_2=2t$, and that the values of the continuum threshold
$\omega_0$ in the sum rules for two-- and three--point functions
coincide. We end up with the set of sum rules\footnote{In this paper
we conform to the standard duality region, which is a square in the
$(s,s')$ plane. Its size is fixed by the continuum threshold in the
corresponding two--point sum rule. Our sum rules are not sensitive to
the detailed shape of the duality region near the diagonal, which is
different from the situation considered in \cite{BS93}.}
\begin{eqnarray}
\label{SR}
\frac{1}{2}F^2(\mu) e^{-\bar\Lambda/t} & = &
\int\limits_0^{\omega_0}\!\! ds\,  e^{-s/t} \rho_{\Pi}(s),
\nonumber\\
\frac{1}{4}F^2(\mu) K e^{-\bar\Lambda/t} & = &
\int\limits_0^{\omega_0}\!\! ds\!\int\limits_0^{\omega_0}\!\!
ds'\,e^{-(s+s')/(2t)} \rho_K(s,s'),\nonumber\\
\frac{1}{4}F^2(\mu) \Sigma(\mu) e^{-\bar\Lambda/t} & = &
 \int\limits_0^{\omega_0}\!\! ds\!\int\limits_0^{\omega_0}\!\!
ds'\,e^{-(s+s')/(2t)} \rho_\Sigma (s,s').
\end{eqnarray}
The normalization scale $\mu$ should be taken to be of order of the
typical Borel parameter. The quantities $\bar\Lambda$ and
$\omega_0$, as well as the ``working region'' in the Borel parameter
$t$ are determined from the two--point sum rule for $F^2$, and
the sum rules for $K$ and $\Sigma$ do not contain
free parameters. In practice, it turns out to be convenient to
consider the ratio of three--point and two--point sum rules, which
suppresses significantly all the sources of uncertainties.
Adding a factor 4 (cf.\ (\ref{splitting})) and taking into account
the scale dependence, we obtain the following sum
rule for the mass splitting between vector and pseudoscalar mesons:
\begin{equation}
m^2_V -m^2_P = 8
 \left(\frac{\alpha_s(m_Q)}{\alpha_s(2t)}\right)^{3/\beta_0}
\,\,\frac{\displaystyle
 \int\limits_0^{\omega_0}\!\! ds\!\int\limits_0^{\omega_0}\!\!
ds'\,e^{-(s+s')/(2t)}\rho_\Sigma (s,s')}{\displaystyle
\int\limits_0^{\omega_0}\!\! ds\,  e^{-s/t} \rho_{\Pi}(s)}\,.
\label{SRsplit}
\end{equation}
In the spectral functions, the renormalization scale is $\mu=2t$.
A similar ratio determines the kinetic energy $K$ of the heavy quark:
\begin{equation}
K = \frac{\displaystyle  2 \int\limits_0^{\omega_0}\!\! ds\!\!
\int\limits_0^{\omega_0}\!\! ds'\,e^{-(s+s')/(2t)}\rho_K (s,s')}
{\displaystyle\int\limits_0^{\omega_0}\!\! ds\,
e^{-s/t}\rho_{\Pi}(s)}\,.\label{SRkin-1}
\end{equation}
Note that the explicit dependence on $\bar\Lambda$ has cancelled out.
It is, however, present implicitly
since the values of $\omega_0$ and $\bar\Lambda$ are correlated.

In the numerical calculations we use the following standard
values of the vacuum condensates (at the normalization point
1 GeV):
\begin{eqnarray}
  \langle\bar q q\rangle &=& -(240\, {\rm MeV})^3,
\nonumber\\
\left\langle\frac{\alpha_s}{\pi} G^2 \right\rangle
&=& 0.012\, {\rm GeV}^4,
\nonumber\\
\langle \bar q \sigma gG q\rangle & = & (0.8\, {\rm GeV}^2)
\langle\bar q q\rangle.
\end{eqnarray}
In this section we use the one--loop expression for the
running coupling with  $\Lambda_{\text{QCD}}=
150\,{\rm MeV}$ and four active
flavors. The corresponding values of the coupling are
$\alpha_s(1 {\rm GeV}) =0.40$ and $\alpha_s(m_B) =0.21$.

The results are shown in Fig.~\ref{fig:5}. For definiteness, we
give values of the decay constant and the chromomagnetic splitting
normalized at the scale of the B--meson mass.
The sum rule for the two--point function is  most stable for
the value $\bar\Lambda = (0.4-0.5)\,\text{GeV}$ and $\omega_0 =
(1.0-1.2)\,{\rm GeV}$, see Fig.~\ref{fig:5}(a). The stability
region starts already at values of the Borel parameter of order
\mbox{0.3 GeV}, and stretches practically to $t\rightarrow\infty$.
It is known, however, that stability at large values of the Borel
parameter is not informative, since in this region the sum rule is
very strongly affected by the continuum model. The usual criterium
that both the higher order power corrections and the contribution of
the continuum should not be very large (say, less than (30--50)\%\/),
restricts the working region considerably. In the particular case of
the decay constant $F(\mu)$ one usually chooses $0.3\,{\rm GeV}<t<
 1\, {\rm GeV}$ \cite{BBBD}.
In the case of the three--point functions, the sum rules are
especially strongly affected by the subtraction of the
continuum owing to the high dimension of the spectral densities.
Thus, it is especially difficult in this case to ensure that
the continuum contribution is sufficiently small, and this
requirement forces one to work in a rather narrow
region of the Borel parameter, close to the lowest possible value
$t\sim 0.3\, {\rm GeV}$. This strategy is backed up by considerable
experience of QCD sum rule calculations of higher--twist operators
built of light quarks, see e.g. \cite{BKY,BK}.
In this paper we take the working window in the Borel parameter
for the three--point functions (\ref{CFK}) and (\ref{CFS}) to be
$0.3\,{\rm GeV}< t< 0.6\, {\rm GeV}$. We emphasize that
this region should be fixed by considering the two--point function,
and does not necessarily coincide with the stability plateau for
three--point sum rules. The remaining instability should be considered
as a part of the errors involved in the calculation.

In Fig.~\ref{fig:5}(b) we plot the right-hand side of the sum rule for
the heavy quark kinetic energy, Eq.~(\ref{SRkin-1}), and in
Fig.~\ref{fig:5}(c) the right-hand side of the sum rule (\ref{SR}),
as a function of the Borel parameter for two different values of the
continuum threshold $\omega_0$ =  1.0 {\rm GeV} and $\omega_0$ =
1.2 {\rm GeV}. The sensitivity of the sum rules to the change of
$\omega_0$ and of the Borel parameter (within the working region
which is shown as shaded area) gives a conservative estimate for
the accuracy of the results. We end up with the values
\begin{eqnarray}
(m^2_V -m^2_P)^{(\mu=m_B)} & = & (0.46\pm 0.14)\,
{\rm GeV}^2\,,\label{sigma}\\
 K^{(\text{LO})} & = & -(0.54\pm 0.09)\,{\rm GeV}^2\,.\label{K1}
\end{eqnarray}
The given value of the mass splitting between vector and
pseudoscalar mesons is our final result whereas the analysis of $K$
will be extended to next--to--leading order in the next section (the
superscript ``LO'' stands for ``leading--order''
result). The experimental values for the mass splitting in beautiful
and charmed mesons are \cite{PDT}
\begin{eqnarray}
   m_{B^*}^2 - m_B^2 &\simeq& 0.48 \,{\rm GeV}^2 \,,\nonumber\\
   m_{D^*}^2 - m_D^2 &\simeq& 0.56 \,{\rm GeV}^2 \,.
\label{splitdata}
\end{eqnarray}
These values are close to each other which indicates the smallness
of higher order corrections. The measured splittings agree quite well
with our result in (\ref{sigma}).

\section{The order $\alpha_{\lowercase{s}}$ corrections to the
kinetic energy}

In this section we calculate the $O(\alpha_s)$ corrections to the
sum rule for the kinetic energy. This calculation is
laborious,  but worthwhile, since in the case of the coupling $F$
the radiative corrections proved to be very large, see
\cite{BBBD,BG92}. Following \cite{BBBD}, we introduce the
heavy--light current $\hat{J}_\Gamma$, which is renormalization
group invariant to two--loop accuracy (cf.\ (\ref{running})),
\begin{equation}
\hat{J}_\Gamma = (\bar q \Gamma h_v)^{(\mu)}
\alpha_s(\mu)^{-\gamma_0/(2\beta_0)}
\left( 1- \delta\,\frac {\alpha_s(\mu)}{\pi}\right),
\label{invcurrent}
\end{equation}
with
\begin{equation}
\delta = \frac{\gamma_0}{8\beta_0}
 \left( \frac{\gamma_1}{\gamma_0}- \frac{\beta_1}{\beta_0}\right)
\approx -0.23\,.
\end{equation}
The corresponding invariant coupling $\hat F$ is defined as
(cf.\ (\ref{fhql})):
\begin{equation}\label{eq:Fscale}
\hat{F} = F(\mu)\,\alpha_s(\mu)^{-\gamma_0/(2\beta_0)}\left(1-\delta\,
\frac{\alpha_s(\mu)}{\pi}\right).
\end{equation}
The leading-- and next--to--leading order anomalous dimensions of
$\hat{J}_{\Gamma}$, $\gamma_0$ and $\gamma_1$, as well as the
coefficients of the $\beta$--function are given in (\ref{anomdim}).

The two--point correlation function of the invariant currents
$\hat{J}_\Gamma$ can be expressed in terms of invariant
quantities and the strong coupling $\alpha_s(\mu)$ at the scale of the
external momentum, $\mu = -2\omega$. A calculation of radiative
corrections to the two--point function (\ref{CF2-def}) leads to the
following sum rule for the invariant coupling $\hat F$ \cite{BBBD}:
\begin{eqnarray}
\lefteqn{\hat F^2 e^{-\bar\Lambda/t}=}\nonumber\\
&=& \frac{3}{\pi^2}\,
\alpha_s(2t)^{-\gamma_0/\beta_0}
\int\limits_0^{\omega_0} ds\,s^2\,e^{-s/t}
\left\{
1+ \frac {\alpha_s(2t)}{\pi}
\left(\frac{17}{3} +\frac{4}{9}\pi^2 -2\delta +
\frac{\gamma_0}{2}\ln\frac{s}{t}
\right)\right\}
\nonumber\\
&&\mbox{}-
\left\{1+\frac {\alpha_s(2t)}{\pi}
\left(2-\frac{\Delta\gamma_1}{8\beta_0}\right)\right\}{\cal O}_3+
\frac{{\cal O}_5}{16 t^2}\,\alpha_s(2t)^{(\gamma^{(5)}-2\gamma_0)/
(2\beta_0)}\,,\label{SRFB}
\end{eqnarray}
which is valid to two--loop accuracy.
Here we have introduced the scale--invariant condensates
\begin{eqnarray}
 {\cal O}_3& = &\langle \bar q q \rangle (\mu)
\alpha_s(\mu)^{-\gamma_0^{(3)}/(2\beta_0)}
\left\{1-\frac {\alpha_s(\mu)}{\pi}\frac{\gamma_0^{(3)}}{8\beta_0}
 \left( \frac{\gamma_1^{(3)}}{\gamma_0^{(3)}}-
  \frac{\beta_1}{\beta_0}\right)
\right\}  \,,
\nonumber\\
{\cal O}_4 & = & \left\langle\frac{\alpha_s}{\pi}G^2\right\rangle\,
\left[1 + O(\alpha_s)\right]\,,\nonumber\\
 {\cal O}_5& = &\langle \bar q \sigma g G q \rangle (\mu)
\alpha_s(\mu)^{-\gamma_0^{(5)}/(2\beta_0)}
\left[1 + O(\alpha_s)\right]\,.
\end{eqnarray}
The leading--order anomalous dimensions are $\gamma_0^{(3)} = 2
\gamma_0 = -8$, $\gamma_0^{(5)} = -4/3$, the next--to--leading order
anomalous dimension of the quark condensate equals $\gamma_1^{(3)} =
-404/9 + 40n_f/9$. As a shorthand, we use $\Delta\gamma_1 =
2\gamma_1-\gamma_1^{(3)}= 704/9-112\pi^2/27$. Apart
{}from an overall scaling factor $2\alpha_s(2t)^{-\gamma_0/(2\beta_0)}$,
the sum rule for $\hat F$ in (\ref{SRFB}) differs from the one given
in the last section by terms of $O(\alpha_s)$. As for
the determination of $\Sigma$ and $K$ to one--loop accuracy, we
neglected these terms and consistently used the
leading--order spectral densities (\ref{spectr}) in the sum rules
(\ref{SRsplit}) and (\ref{SRkin-1}).

Our aim now is to calculate the double spectral function $\rho_K$
defined in Eq.~(\ref{disp}) to next--to--leading order. The
calculation of the perturbative two--loop corrections to the diagram
Fig.~\ref{fig:2}(a) is the most difficult task. The necessary
techniques are explained in App.~\ref{app:A}, while the contributions
of the individual diagrams in Fig.~\ref{fig:2} are collected in
App.~\ref{app:B}. The result reads:
\begin{eqnarray}
  \rho_K^{\text{pert}}(s,s')
& = &  \left(-\frac{3}{4\pi^2}\right)\, s^4\,
\delta(s-s')\, \left\{ 1 + \frac{\alpha_s(\mu)}{\pi} \left(
\frac{41}{9} + \frac{4}{9}\pi^2 + \frac{\gamma_0}{2}\,
 \ln \frac{2s}{\mu}\right)\right\}\nonumber\\
& & {} -\frac{\alpha_s(\mu)}
{\pi}\,\frac{1}{2\pi^2} \,(s+s')\left(
s^2 + s'^2 -  |s^2 -s'^2| \right).
\label{pert-2loop}
\end{eqnarray}
In Feynman gauge, the spectral densities of individual
diagrams contain terms like $d/ds\,\ln |s-s'|$ which remind of the
cancellation of infra--red divergences between particular
discontinuities, corresponding to real and virtual gluon emission.
In the sum given above, however, they cancel.

The contribution of the quark condensate
is easier to calculate. The relevant diagrams are collected
in Fig.~\ref{fig:3} and their spectral density is
\begin{eqnarray}
\rho_K^{(3)}(s,s')&=&\frac{1}{3}\,\frac{\alpha_s}{\pi}
\langle \bar q q \rangle
\left\{s\delta(s') +s'\delta(s) +(s+s')\delta(s-s')\right\}.
\label{K3}
\end{eqnarray}
Finally, we take into account the contribution of the gluon
condensate, see the diagrams in Fig.~\ref{fig:4}. The corresponding
contribution to the spectral density equals
\begin{eqnarray}
\rho_K^{(4)}(s,s')&=&-\frac{1}{16} \left\langle \frac{\alpha_s}{\pi}
G^2 \right \rangle \delta(s-s')\,.
\label{K4}
\end{eqnarray}
The leading tree--level contribution of the mixed condensate is given
in (\ref{spectr}).
We neglect the contribution of the four--quark condensate which is
tiny in the case of $\hat F$ and of semileptonic form factors
\cite{BBD}.

The results given above correspond to the calculation of the Feynman
diagrams shown in Figs.~\ref{fig:2}, 3, and 4. To obtain the
correlation function involving invariant currents, one has to multiply
the spectral densities Eqs.~(\ref{pert-2loop}), (\ref{K3}), and
(\ref{K4}) by the corresponding power of the coupling. Then, after the
Borel transformation, renormalization group improvement reduces to the
substitution of the scale $\mu$ by $2t$, twice the Borel parameter
(see \cite{BBBD} for details). Combining all terms, we end up with the
sum rule
\begin{eqnarray}
\hat{F}^2Ke^{-\bar{\Lambda}/t} & = & \int\limits_0^{\omega_0}\!\!
ds\!\!\int\limits_0^{\omega_0}\!\! ds'\,e^{-(s+s')/(2t)}\left(
\alpha_s(2t)^{-\gamma_0/\beta_0}\left[
-\frac{3}{\pi^2}\, s^4\, \delta(s-s')\, \left\{ 1 +
\frac{\alpha_s(2t)}{\pi} \left(\frac{41}{9} + \frac{4}{9}\pi^2
\right.\right.\right.\right. \nonumber\\
& & \left.\left.\left.- 2\delta+ \frac{\gamma_0}{2} \ln \frac{s}{t}
\right)
\right\} -\frac{\alpha_s(2t)}{\pi}\,\frac{2}{\pi^2} \,(s+s')\left(
s^2 + s'^2 -  |s^2 -s'^2| \right)\right]+ \frac{4}{3}\,
\frac{\alpha_s}{\pi}\,{\cal O}_3 \nonumber\\
& &  \left.\left.{}\times\left\{s\delta(s') +s'\delta(s)
\vphantom{\frac{\Delta\gamma_1}{8\beta_0}} + (s+s')\delta(s-s')
\right\}-\frac{1}{4}\,\alpha_s(2t)^{-\gamma_0/\beta_0}
\,{\cal O}_4\, \delta(s-s')\right.\right.
\nonumber\\ & & \left. {} + \frac{3}{8}\,
\alpha_s(2t)^{(\gamma^{(5)}-2\gamma_0)/(2\beta_0)}\,{\cal
O}_5\,\delta(s)\,\delta(s')\right)\,.\label{eq:rhoK2L}
\end{eqnarray}
In both the sum rules (\ref{SRFB}) and (\ref{eq:rhoK2L}) the
continuum contribution is subtracted also from the imaginary part of
the running coupling $\alpha_s(-2\omega)$. The term $\ln s/t$ comes
{}from the expansion of $\{\alpha_s(2s)/\alpha_s(2t)\}^{-\gamma_0/
\beta_0}$ to first order \cite{BBBD}.

In the numerical evaluation of (\ref{eq:rhoK2L}) we use the values of
the condensates as given in the last section and the two--loop formula
for the running coupling,
\begin{equation}
\alpha_s(\mu) = \frac{4\pi}{2\beta_0\ln(\mu/
\Lambda^{(n_f)}_{\overline{\text{MS}}})}\,
\left\{ 1 -
\frac{\beta_1}{\beta_0}\,\frac{\ln(2\ln(\mu/
\Lambda^{(n_f)}_{\overline{\text{MS}}}))}{2 \ln(\mu/
\Lambda^{(n_f)}_{\overline{\text{MS}}})} \right\},
\end{equation}
with $\Lambda_{\overline{\text{MS}}}^{(4)} = 260\,\text{MeV}$ for four
running quark flavors \cite{PDT}. As in the analysis of the
leading--order sum rule
(\ref{SRkin-1}), we vary the Borel parameter $t$ in the range
$0.3\,\text{GeV} < t < 0.6\,\text{GeV}$ and the continuum
threshold in the range $1\,\text{GeV}< \omega_0<
1.2\,\text{GeV}$. In Fig.~\ref{fig:6}(a) we show the coupling
$\hat{F}^2$, calculated according to (\ref{SRFB}) and scaled up to
$\mu=m_B$ according to (\ref{eq:Fscale}). Figure~\ref{fig:6}(b)
contains the kinetic energy $K$, calculated by taking the ratio of
(\ref{eq:rhoK2L})
and (\ref{SRFB}). Both quantities are plotted as functions of the
Borel parameter $t$ for $\omega_0 = 1\,\text{GeV}$
($\bar\Lambda = 0.4\,\text{GeV}$)
(solid lines) and $\omega_0 = 1.2\,\text{GeV}$ ($\bar\Lambda = 0.5
\,\text{GeV}$) (dashed lines). The working region is indicated by the
shaded areas. As can be seen when comparing
the two figures with each other, $K$ is less sensitive to the value of
$\omega_0$ than $\hat{F}^2$, and we get the final result
\begin{equation}\label{eq:result2L}
K = -(0.60\pm 0.10)\,\text{GeV}^2.
\end{equation}
The errors include the dependence of the sum rule (\ref{eq:rhoK2L})
on variations of the Borel parameter $t$ and the continuum threshold
$\omega_0$. In the range of parameters considered, the
non--perturbative terms contribute approximately $25\%$.

The value (\ref{eq:result2L}) is very close to
the leading order result (\ref{K1}).
 The two--loop radiative
corrections in the sum rules are large and change the value of
the coupling $F^2(m_B)$ by a factor two, cf.\ Figs.~\ref{fig:5}(a)
and \ref{fig:6}(a). Nevertheless, their net effect on the value of $K$
is small due to a strong cancellation in the ratio of (\ref{eq:rhoK2L})
and (\ref{SRFB}). It is interesting to note that the Coulombic
radiative corrections, which contain an extra factor $\pi^2$, are
identical for the two-- and the three--point sum rules. A similar
effect was observed in the study of the Isgur--Wise function
in \cite{BBG}.

\section{Discussion}

We have derived sum rules for three--point correlation functions
in the heavy quark effective theory, from which we obtain estimates
for the matrix elements of the chromomagnetic interaction and the
kinetic energy operator over heavy--light mesons. Our final results
read
\begin{eqnarray}
 \frac{\langle M(v)|{\cal K}|M(v)\rangle}{
\langle M(v)|\bar h_v h_v|M(v)\rangle} &=&
 -(0.60\pm 0.10)\,\text{GeV}^2
\nonumber\\
 \frac{\langle M(v)|{\cal S}|M(v)\rangle^{\mu=m_B}
 }{\langle M(v)|\bar h_v h_v|M(v)\rangle} &=&
 \frac{d_M}{4}[(0.46\pm 0.14)\,\text{GeV}^2 ]\,.
\label{final}
\end{eqnarray}
Note that the value of kinetic energy is renormalization group
invariant, while the chromomagnetic interaction depends on the
heavy quark mass scale. The latter dependence is actually
significant. The given value (in square brackets)
at  $\mu = m_B \simeq 5.3 \,\text{GeV}$
corresponds to  $(0.60\pm 0.18)\,\text{GeV}^2$ at the lower
scale $\mu = 1\, \text{GeV}$.

Our result for the kinetic energy is two times lower than the
value obtained in \cite{N92} from the expansion of two--point
sum rules in QCD in the heavy quark limit. The approach of the
present paper should be much more accurate, since  the consideration
of three--point functions allows one to suppress contaminating
contributions of non--diagonal transitions. Still, this disagreement
is disturbing, since QCD sum rules normally have a (20--30)\% accuracy.
The large value of the kinetic energy obtained
in  \cite{N92} may be an artifact of the used procedure with a
redefinition of the Borel parameter by terms of order $1/m_Q$.
{}From our point of view, no such redefinitions are allowed.

The
corrections to B--meson masses,  corresponding to (\ref{final}), are
\begin{eqnarray}
m_B &=& m_b +\bar\Lambda +\frac{1}{2m_b}[(0.25\pm 0.20)\,\text{GeV}^2]
\,,
\nonumber\\
m_{B^*} &=& m_b +\bar\Lambda +\frac{1}{2m_b}[(0.70\pm 0.20)\,
\text{GeV}^2]\,,
\end{eqnarray}
respectively.  Note that the
$1/m_b$ correction to the  pseudoscalar B--meson
is very small. This gives
strong support to the procedure of the evaluation of the $1/m_Q$
correction to the leptonic decay constant in \cite{BBBD},
where the $1/m_Q$ correction to the pseudoscalar meson mass have been
neglected. For completeness, we quote the result of \cite{BBBD}
\begin{equation}
  \hat F (m_Q) = \hat F \left [ 1- \frac{(0.8-1.1)\text{GeV}}
{m_Q}\right ]
\end{equation}
which agrees to all other QCD sum rule calulations \cite{SE91,NA},
with the only exception of \cite{N92}, where a much larger
correction is claimed.

The phenomenological applications of our results include $1/m_Q$
corrections to semileptonic form factors and inclusive B--decays.
The rather small value of the kinetic energy obtained in this paper
may require the reconsideration of recent estimates of these
corrections in \cite{shifetc}.

\acknowledgements
We are grateful to  H.G.\ Dosch and N.G.\ Uraltsev for
interesting and stimulating discussions.
Our special thanks are due to A.V.\ Radyushkin, who found an error
in the preprint version of this paper.

\appendix
\section{The 1/\lowercase{m}$_{\text{Q}}$ expansion of correlation
functions}\label{app:C}

The aim of this Appendix is to establish a formal connection
between the $1/m_Q$ expansion of correlation functions in QCD,
and the corresponding correlation functions in the HQET.
Let us consider the following two--point correlation
functions of vector and pseudoscalar currents:
\begin{eqnarray}
\Pi_{\mu\nu}(q) &=& i\int d^4x\,e^{iqx}\langle 0 |
T \left\{ \bar q(x) \gamma_\mu b(x) \bar b(0) \gamma_\nu q(0)
 \right\}|0\rangle
\nonumber\\
 &&\mbox{}= \left(-g_{\mu\nu}+\frac{q_\mu q_\nu}{q^2} \right)
\Pi_v(q^2) +\frac{q_\mu q_\nu}{q^2}\Pi_s(q^2)\,,
\nonumber\\
\Pi_5(q^2) &=& i\int d^4x\,e^{iqx}\langle 0 |
 T \left\{ \bar q(x)i\gamma_5 b(x) \bar b(0)i\gamma_5 q(0)
  \right\}|0\rangle\,.
\label{twopoint}
\end{eqnarray}
The correlation functions $\Pi_v(q^2)$ and $\Pi_5(q^2)$ have poles
at $q^2 = m_{V}^2$ and $q^2 = m_P^2$, respectively, and
the contribution of the ground state mesons in the vector chanel, $V$,
and in the pseudoscalar chanel, $P$, equals
 \begin{eqnarray}
 \Pi_v(q^2) &=& \frac{f_V^2 m_V^2}{m_V^2-q^2} +\ldots,
\nonumber\\
 \Pi_5(q^2) &=& \frac{f_P^2 m_P^4}{m_Q^2(m_P^2-q^2)} +\ldots
\label{pole}
 \end{eqnarray}
where the couplings $f_P$ and $f_V$ are defined as in
(\ref{couplings}).
Our objective is to make a systematic expansion of the correlation
functions in (\ref{twopoint}) near the poles (\ref{pole})
in powers of the large quark mass $m_Q$.  Each correlation
function has symmetric pairs of poles at $q_0 = \pm m_M$, which
correspond in an obvious way to contributions of particles and
antiparticles. For definiteness, we consider the expansion near
the particle--type discontinuity, and to this end define a new
variable $\omega$ by
\begin{equation}
    q_\mu = v_\mu(m_Q+\omega).
\label{omega}
\end{equation}
The physical couplings and masses of the mesons are expanded as
\begin{eqnarray}
   f_M\sqrt{m_M}& =& F_M+ \delta F_M ,
\nonumber\\
   m_M &=& m_Q +\bar \Lambda + \delta m_M ,
\label{expand}
 \end{eqnarray}
where $\delta F_M$ and $\delta m_M$ are of order $1/m_Q$. We
anticipate here that the mass splitting between pseudoscalar, $M=P$,
and vector mesons, $M=V$, is an $1/m_Q$ effect. Note that as a quite
general feature of QCD, the extraction of asymptotic
behavior (in our case in the heavy quark mass)
introduces divergences. The separation
of the asymptotic value of the coupling $F$ tacitly assumes
some regularization and the
summation of logarithms of the heavy quark mass, which is made
explicit using the formalism
of the heavy quark effective theory.

Putting together Eqs.~(\ref{pole}), (\ref{omega}), and (\ref{expand}),
we obtain, e.g. for the vector correlation function:
\begin{equation}
 \frac{f_{B^*}^2 m_{B^*}^2}{m_{B^*}^2-q^2} =
 \frac{F^2}{2\Delta} -\frac{2 \delta m F^2}{(2\Delta)^2}
  +\frac{2 F \delta F}{2\Delta} +\mbox{non--singular terms,}
\end{equation}
where
\begin{equation}
 \Delta = \bar \Lambda -\omega.
\end{equation}
The expansion of the correlation functions in (\ref{twopoint}) is
not so immediate. Using the heavy quark expansion of currents,
Eq.~(\ref{matching}), and the $1/m_Q$ expansion of the
Lagrangian, Eq.~ (\ref{Lagrangian}), we obtain for the vector
correlation function
\begin{eqnarray}
\lefteqn{    i\int d^4x\,e^{i(v\cdot x)(m_Q+\omega)}\langle 0 |
T \left\{ \bar q(x) \gamma_\mu Q(x) \bar Q(0) \gamma_\nu q(0)
 \right\}|0\rangle =}
\nonumber\\
&&\mbox{}=\left( 1-\frac{8}{3}\frac{\alpha_s(m_Q)}{\pi}\right)
i\int d^4x\,e^{i\omega(v\cdot x)}\langle 0 |  T \left\{ \bar q(x)
 \gamma_\mu^\perp h_v(x) \bar h_v(0) \gamma_\nu^\perp q(0)
 \right\}|0\rangle^{(\mu=m_Q)}
\nonumber\\
&&\mbox{}+\frac{1}{2m_Q}
i\int d^4x\,e^{i\omega (v\cdot x)}\langle 0 |  T \left\{ \bar q(x)
 \gamma_\mu^\perp (i\not\!\! D^\perp)
  h_v(x) \bar h_v(0) \gamma_\nu^\perp q(0)
 \right\}|0\rangle^{(\mu=m_Q)}
\nonumber\\
&&\mbox{}+\frac{1}{2m_Q}
i\int d^4x\,e^{i\omega (v\cdot x)}\langle 0 |  T \left\{ \bar q(x)
 \gamma_\mu^\perp  h_v(x) \bar h_v(0)
  (-i\stackrel{\leftarrow}{\not\!\!D}^\perp)
  \gamma_\nu^\perp q(0)
 \right\}|0\rangle^{(\mu=m_Q)}
\nonumber\\
&&\mbox{}+\frac{1}{2m_Q}
i^2\int d^4x\,d^4y\,e^{i\omega(v\cdot x)}\langle 0 |
 T \left\{ \bar q(x)
 \gamma_\mu^\perp  h_v(x)  {\cal K}(y) \bar h_v(0)
  \gamma_\nu^\perp q(0)
 \right\}|0\rangle^{(\mu=m_Q)}
\nonumber\\
&&\mbox{}+\frac{1}{2m_Q}
i^2\int d^4x\,d^4y\,e^{i\omega (v\cdot x)}
\langle 0 |  T \left\{ \bar q(x)
 \gamma_\mu^\perp  h_v(x)  {\cal S}(y) \bar h_v(0)
  \gamma_\nu^\perp q(0)
 \right\}|0\rangle^{(\mu=m_Q)}
\nonumber\\
 &&\mbox{} +\mbox{longitudinal part} +
 O\left(\frac{1}{2m_Q} \frac{\alpha_s(m_Q)}{\pi}\right) +O(1/m_Q^2).
\label{CF-expand}
\end{eqnarray}
 Most of the correlation functions in (\ref{CF-expand})
 have been introduced already in the main text. In addition to
them, we encounter one more invariant function,
\begin{equation}
i\int d^4x\,e^{i\omega vx}\langle 0 |  T \left\{ \bar q(x)
 \Gamma_1(i D^\perp_\alpha)
  h_v(x) \bar h_v(0) \Gamma_2 q(0)
 \right\}|0\rangle
   = -\frac{1}{2} \mbox{Tr}\{\Gamma_1 P_+ \Gamma_2 \} \Pi_1(\omega),
\end{equation}
which can be related, however, to the invariant function
$\Pi(\omega)$  in (\ref{CF2-def}), choosing $\Gamma_1 = \gamma_\alpha$,
integrating by parts, and using the equations
of motion  for the effective heavy quark field
$ (vD)h_+ =0$ and for the
light antiquark $\bar q
(-i\stackrel{\leftarrow}{\not\!\!D})= 0$.
One obtains \cite{N92}:
\begin{equation}
\Pi_1(\omega)= -\frac{1}{3}\, \omega\, \Pi(\omega).
\end{equation}
Combining everything we arrive at
\begin{equation}\label{expvec}
\Pi_v(q^2 = (m_Q+\omega)^2) =
 \left(1-\frac{8}{3}\frac{\alpha_s(m_Q)}{\pi}\right)
 \Pi(\omega) +\frac{\omega}{3m_Q}\Pi(\omega)
+\frac{1}{m_Q} T_K(\omega,\omega)
 -\frac{1}{m_Q}T_\Sigma(\omega,\omega).
\end{equation}
The expansion of the correlation function of pseudoscalar currents
is obtained by similar manipulations. The result reads
\begin{equation}\label{exppseudo}
\Pi_5(q^2 = (m_Q+\omega)^2) =
 \left(1-\frac{4}{3}\frac{\alpha_s(m_Q)}{\pi}\right)
 \Pi(\omega) +\frac{\omega}{m_Q}\Pi(\omega)
+\frac{1}{m_Q} T_K(\omega,\omega)
+\frac{3}{m_Q}T_\Sigma(\omega,\omega).
\end{equation}
Note that these relations are valid up to constant terms only,
which come from distances $\sim 1/m_Q$ and are lost in the
heavy quark expansion. Using the operator product expansion results
for the vector and pseudoscalar correlation functions in QCD,
it is easy to verify that the
answers for the contributions to
the correlation functions in HQET of the quark and the mixed quark-gluon
condensate given in the main text indeed satisfy the general relations
in (\ref{exppseudo}) and (\ref{expvec}). For contributions of
the pertubation theory and the gluon condensate this check is not so
immediate, since in the process of calculation
we have discarded contributions which vanish after the double Borel
transformation.

 The last step is to insert a complete set of meson states in the
correlation functions and to separate the contribution of
the lowest energy. Note that at $\omega=\omega'$
the three--point correlation functions $T_K(\omega,\omega')$ and
$T_\Sigma(\omega,\omega')$ contain
contributions with both a double pole and a single pole
at $\omega =\bar \Lambda $:
\begin{eqnarray}
T_K(\omega,\omega)&=& \frac{F^2(\mu)K}{[2(\bar\Lambda -\omega)]^2}
+ \frac{2F^2(\mu)G_K}{2(\bar\Lambda -\omega)}
+\mbox{non--singular terms} ,
\nonumber\\
T_\Sigma(\omega,\omega)&=
& \frac{F^2(\mu)\Sigma}{[2(\bar\Lambda -\omega)]^2}
+ \frac{2F^2(\mu)G_\Sigma}{2(\bar\Lambda -\omega)}
+\mbox{non--singular terms} ,
\end{eqnarray}
where $K$ and $\Sigma$ are matrix elements of the kinetic energy
and the chromomagnetic interaction operator
over the heavy--light meson in the effective theory.
The quantities $G_K$ and $G_\Sigma$ are defined as
\begin{eqnarray}
\langle 0|\, i\!\int\! dx\, {\cal K}(x) \bar h_v(0) \Gamma q(0)
|M(v)\rangle &=& \frac{1}{2}F(\mu)G_K(\mu)
\mbox{Tr}\{\Gamma {\cal M}(v)\}\,,
\nonumber\\
\langle 0|\, i\!\int\! dx\, {\cal S}(x) \bar h_v(0) \Gamma q(0)
|M(v)\rangle &=& \frac{1}{2}F(\mu)G_\Sigma(\mu)
2 d_M \mbox{Tr}\{\Gamma {\cal M}(v)\}\,,
\end{eqnarray}
where it is implied that the matrix elmements of the two--point
correlation functions on the left--hand side include summation
over all intermediate states {\em except} for the lowest--lying
one. For example,
\begin{eqnarray}
\lefteqn{\langle 0| i\!\int dx \,{\cal K}(x) \bar h_v(0) \Gamma q(0)
|M(v)\rangle = }
\nonumber\\
&&\mbox{}= \sum_{M'\neq M}
\langle 0|{\cal K}|M'(v)\rangle\,
\frac{1}{2m_B(\Lambda'-\bar \Lambda)}
\, \langle M'(v)| \bar h_v \Gamma q |M(v)\rangle
\end{eqnarray}
where $\Lambda' = m_{M'}-m_Q$. Up to this remark, our
formul\ae\/ coincide with the ones derived in \cite{N92}.

Collecting everything and comparing to the expansion of the
relativistic expression in Eq.~(\ref{expand}), we obtain
\begin{eqnarray}
\delta m &=& -\frac{1}{2m_Q}\left[K+ d_M \Sigma\right],
\nonumber\\
\frac{\delta F}{F}&=&-\frac{\bar\Lambda}{6 m_Q} d_M +
\frac{1}{2m_Q} G_K + \frac{d_M}{m_Q} G_\Sigma ,
\nonumber\\
2m_P(m_V-m_P)&=&m^2_V-m_P^2 = 4\Sigma,
\label{correction}
\end{eqnarray}
which is the desired result. The relations in
(\ref{correction}) can of course be derived in a variety of ways
and are by no means new. Our derivation
is suited best for the application of the sum rule
technique to the evaluation of relevant parameters.
\section{Calculation of spectral densities to two--loop
accuracy}\label{app:A}

In this appendix we give all the formul\ae\/ necessary to
calculate the spectral densities of the correlation functions
(\ref{CFK}) and (\ref{CFS}) to two--loop accuracy.
To this end, we use a technique proposed by \cite{NR82} which amounts
to calculate double spectral densities by a repeated application of
the Borel transformation to the amplitude itself. Let $A$ be some
amplitude depending on the virtualities $\omega$ and $\omega'$, which
we want to  write as a dispersion relation (subtracted if necessary)
in both variables:
\begin{equation}
A(\omega,\omega') = \int\!\! ds\,
ds'\,\frac{\rho(s,s')}{(s-\omega) (s'-\omega)} + {\rm
subtractions}.
\end{equation}
Let us assume that the amplitude $A(\omega,\omega')$ is calculated,
e.g.\ in the from of a Feynman parameter integral. Then, applying a
Borel transformation in $\omega$ and $\omega'$, we obtain, as
the second step:
\begin{equation}
\hat{B}_\omega(t)\,\hat{B}_{\omega'} (t')\,A(\omega,\omega') =
\hat{A}(t,t') = \frac{1}{tt'}\!\int\!\! ds\, ds'\,
\rho(s,s')\,e^{-s/t - s'/t'}.
\end{equation}
Finally, we apply a Borel transformation in $\tau=1/t$ and
$\tau'=1/t'$ to get
\begin{equation}
\hat{B}_\tau\left(1/s \right)\,
\hat{B}_{\tau'}\left(1/s' \right)\, \frac{1}{\tau\tau'}\,
\hat{A}\left(1/\tau,1/\tau' \right)  =  s s' \rho(s,s'),
\end{equation}
which is (apart from a trivial factor) just the desired spectral
density.

Here we made repeated use of the formula (with $\omega_E= -\omega$)
\begin{equation}\label{eq:Borsimpel}
\hat{B}_\omega(t)\,e^{-\alpha \omega_E} = \delta(1-\alpha t).
\end{equation}
We spend one subsection to derive necessary formul\ae\/ for each of
the above three steps, and finally illustrate the whole procedure
 by the sample calculation of a one and a two--loop diagram.

\subsection{Reduction of Loop--Integrals to One--Parameter Integrals}

The general one--loop integral is defined as
\begin{equation}\label{eq:master1L}
I_1(\alpha,p,q) = \int\!\! \frac{d^D\!k}{(2\pi)^D}\,
\frac{1}{[k^2]^\alpha (\omega+vk)^p (\omega'+vk)^q}.
\end{equation}
Using the parametrisation
\begin{equation}
\frac{1}{A^\alpha P^p} =
\frac{\Gamma(\alpha+p)}{\Gamma(\alpha)\Gamma(p)}\!\!
\int\limits_0^\infty\!\! dy \, \frac{y^{p-1}}{(A+yP)^{\alpha+p}}
\end{equation}
we find
\begin{equation}
I_1(\alpha,p,q) = (-1)^{\alpha+p+q}\,\frac{i}{4^\alpha\pi^{D/2}}
\,J(\alpha,p,q)\,F(q-1,p-1,2\alpha+p+q-D)
\end{equation}
with
\begin{eqnarray}
J(\alpha,p,q) & = & \frac{\Gamma(2\alpha+p+q-D)
\Gamma(D/2-\alpha)}{\Gamma(\alpha)\Gamma(p)\Gamma(q)},\\
F(a,b,c) & = & \int\limits_0^1\!\! dz\,z^a\,(1-z)^b\,[(\omega'_E z +
\omega_E (1-z)]^{-c},
\end{eqnarray}
where we have already continued to Euclidean space: $\omega_E =
-\omega$, $\omega'_E = -\omega'$. In the limit $q\to 0$ we find
\begin{equation}
I_1(\alpha,p,0) =
(-1)^{\alpha+p}\,\frac{i}{4^a\pi^{D/2}}\,I(\alpha,p)\,
\omega_E^{D-2\alpha-p}
\end{equation}
with
\begin{equation}
I(\alpha,p) =
\frac{\Gamma(2\alpha+p-D)\Gamma(D/2-\alpha)}{\Gamma(\alpha)\Gamma(p)}
\end{equation}
and a corresponding formula for $p\to 0$.

The general two--loop integral can be expressed as
\begin{eqnarray}
\lefteqn{I_2(\alpha,\beta,\gamma,p,q,r,s) =}\nonumber\\
& = & \int\!\! \frac{d^D\! k}{(2\pi)^D}\,\frac{d^D\!
l}{(2\pi)^D}\,\frac{1}{[k^2]^\alpha\,[l^2]^\beta\,[(k-l)^2]^\gamma\,
(\omega+vk)^p\,(\omega+vl)^q\, (\omega'+vk)^r\, (\omega'+vl)^s}.
\end{eqnarray}
Simple cases are
\begin{eqnarray}
\lefteqn{I_2(\alpha,\beta,0,p,q,r,s) = I_1(\alpha,p,r)\,
I_1(\beta,q,s),}\\
\lefteqn{I_2(\alpha,0,\gamma,p,0,r,s) =
\frac{(-1)^{\alpha+\gamma+p+r+s+1}}{4^{\alpha+\gamma}\pi^D}\,
I(\gamma,s)\,J(\gamma,p,r+s+2\alpha-D)}\nonumber\\
& & \times
F(r+s+2\alpha-D-1,p-1,2(\alpha+\gamma-D)+p+r+s),\\
\lefteqn{I_2(\alpha,0,\gamma,0,q,0,s) = \frac{\Gamma(D/2-\alpha)
\Gamma(D/2-\gamma)\Gamma(2(\alpha+\gamma-D)+q+s)}{
\Gamma(\alpha)\Gamma(\gamma)\Gamma(q)\Gamma(s)}}\nonumber\\
& & \times
\frac{(-1)^{\alpha+\gamma+q+s+1}}{4^{\alpha+\gamma}\pi^D}\,
F(s-1,q-1,2(\alpha+\gamma-D)+q+s),\\
\lefteqn{I_2(\alpha,0,\gamma,p,0,0,s) = \frac{(-1)^{\alpha+\gamma+p+s+
1}}{4^{\alpha+\gamma}\pi^D}\,I(\gamma,s)\, J(\alpha,p,s+2\gamma-D)
}\nonumber\\ & & \times F(s+2\gamma-D-1,p-1,2(\alpha+\gamma-D)+p+s),\\
\lefteqn{I_2(\alpha,\beta,\gamma,p,0,r,0) =
\frac{(-1)^{\alpha+\beta+\gamma+p+r+1}}{4^{\alpha+\beta+\gamma}\pi^D}
\,F(r-1,p-1,2(\alpha+\beta+\gamma-D)+p+r)
}\nonumber\\& & \times \,J(\alpha+\beta+\gamma-D/2,p,r)\,
\frac{\Gamma(\beta+\gamma-D/2) \Gamma(D/2-\beta)\Gamma(D/2-\gamma)}{
\Gamma(\beta)\Gamma(\gamma)\Gamma(D-\beta-\gamma)}.\makebox[2cm]{}
\end{eqnarray}
Similar formul\ae\/ can be obtained be exchanging $\omega$ and
$\omega'$ or the loop momenta.
The special case $r=s=0$ is considered in \cite{BG91} where
recursion relations for $I_2$ were derived by using the method of
integration by parts (cf.\ \cite{BG91,CT81}). We employ the same
technique to relate the general $I_2$ to the special cases solved
above and find the following recursion relations:
\begin{eqnarray}
\lefteqn{I_2(\alpha,0,\gamma,p,q,r,s) =}\nonumber\\
& & = \frac{1}{2\gamma+q+s-D}\, \{\,q\,
I_2(\alpha,0,\gamma,p-1,q+1,r,s) +
s\,I_2(\alpha,0,\gamma,p,q,r-1,s+1)\,\},\makebox[0.5cm]{}\\
\lefteqn{I_2(\alpha,0,\gamma,0,q,r,s) =  \frac{1}{2\gamma+s-D}\, \{ \,
q\,I_2(\alpha,0,\gamma,0,q+1,r-1,s)}\nonumber\\
& & {} - q\,I_2(\alpha,0,\gamma,0,q+1,r,s-1) + s\,
I_2(\alpha,0,\gamma,0,q,r-1,s+1)\,\},\\
\lefteqn{I_2(\alpha,\beta,\gamma,0,q,r,0) =
\frac{1}{\beta-\gamma-\frac{\omega'}{\omega}\,(2\beta+\gamma+q-D)}\,
\{\beta\,I_2(\alpha,\beta+1,\gamma-1,0,q,r,0)} \nonumber\\
& & {}-\beta\, I_2(\alpha-1,\beta+1,\gamma,0,q,r,0) -
q I_2(\alpha,\beta,\gamma,0,q+1,r-1,0)\nonumber\\
& & {}+ \gamma \left(1-\frac{\omega'}{\omega}\right)\,
{}[\, I_2(\alpha-1,\beta,\gamma+1,0,q,r,0) -
I_2(\alpha,\beta-1,\gamma+1,0,q,r,0)\, ]\, \},\\
\lefteqn{I_2(\alpha,\beta,\gamma,p,q,r,0) =
\frac{1}{D-2\gamma-\beta-q}\,\{
\beta\,I_2(\alpha,\beta+1,\gamma-1,p,q,r,0)}\nonumber\\
& &{} - \beta\,I_2(\alpha-1,\beta+1,\gamma,p,q,r,0) -
q\,I_2(\alpha,\beta,\gamma,p-1,q+1,r,0)\}.
\end{eqnarray}
Note that one has $I_2(\alpha,\beta,\gamma,p,q,r,s) =
I_2(\beta,\alpha,\gamma,q,p,s,r)$.\\
Finally, we encounter the integrals
\begin{eqnarray}
A & = & \int\!\! \frac{d^Dk\,d^Dl}{(2\pi)^{2D}}\,
\frac{k^2}{l^2(l-k)^2(\omega+vk)(\omega'+vk)(\omega'+vl)},\\
B & = & \int\!\! \frac{d^Dk\,d^Dl}{(2\pi)^{2D}}\,
\frac{k^2}{l^2(l-k)^2(\omega+vk)(\omega+vl)(\omega'+vk)(\omega'+vl)}.
\end{eqnarray}
After a change in variables, $k\to k+l$, and noting that the integral
over $l$ can depend on $v$ only, we find
\begin{eqnarray}
A & = & 2\!\int\!\! \frac{d^Dk\,d^Dl}{(2\pi)^{2D}}\,
\frac{(vl)(vk)}{l^2 k^2(\omega'+vk) (\omega+vk+vl) (\omega'+vk +vl)}\\
& = & 2\omega'_E\,I_2(0,1,1,1,0,0,1)-2i\,
\frac{(-1)^{D/2}}{(4\pi)^{D/2}}
\,\frac{\Gamma(D/2)\Gamma(D/2-1)\Gamma(2-D/2)}{\Gamma(D-1)}\nonumber\\
& & \times\omega_E \,I_1(2-D/2,1,1).
\end{eqnarray}
Similarly, one obtains
\begin{equation}
B = 2\omega_E\, \{ \omega_E'\,I_2(0,1,1,1,1,0,1) - I_2(0,1,1,1,0,1,0) -
\omega_E\, I_2(0,1,1,1,1,1,0)\}.
\end{equation}

\subsection{First Application of the Borel Transformation}

In this appendix we are concerned with the calculation of
\begin{equation}
\hat{K}(p,q;\alpha,\beta,\gamma) = \hat{B}_\omega(t)\,
\hat{B}_{\omega'}(t')\, \omega_E^\alpha\, \omega'^\beta_E\,
F(p,q,\gamma).
\end{equation}
For non--negative integer $\alpha$, $\beta$ we can write
\begin{eqnarray}
\hat{K}(p,q,\alpha,\beta,\gamma) & = & \hat{B}_\omega(t)\,
\hat{B}_{\omega'}(t') \frac{1}{\Gamma(\gamma)}\,\omega_E^\alpha\,
\omega_E'^\beta \int\limits_0^1\!\! dz\, z^p\, (1-z)^q
\int\limits_0^\infty \!\! ds\, s^{\gamma-1}\, e^{-s(\omega_E' z +
\omega_E (1-z))}\nonumber\\
& = & \frac{(-1)^{\alpha+\beta}}{\Gamma(\gamma)}\, t^\alpha\,
t'^\beta\, \frac{d^{\alpha+\beta}}{dt^\alpha\,dt'^\beta}\,
t^{1+\alpha+p-\gamma}\, t'^{1+\beta+q-\gamma}\, (t+t')^{\gamma-p-q-2}.
\end{eqnarray}
Likewise we need $\hat{K}(0,0,\alpha\not\in {\bf N},\beta\in {\bf
N},\gamma)$. In the case $1+\beta+[\alpha]\leq0$, where the
Gaussian bracket [$\alpha$] denotes the integer part of $\alpha$,
one finds
\begin{eqnarray}
\lefteqn{\hat{K}(0,0,\alpha\not\in {\bf N},\beta\in {\bf
N},\gamma)}\nonumber\\
 & = & \frac{(-1)^\beta}{\Gamma(-\alpha)\Gamma(\gamma)}\,t^\alpha
t'^\beta\,\frac{d^\beta}{dt'^\beta}\, \frac{t'^{1+\alpha+
\beta-\gamma}}{(t+t')^{1+\alpha}}
\,\int\limits_{\frac{t}{t+t'}}^1\! dz\,
z^{1+\alpha-\gamma}\, \left( z- \frac{t}{t+t'}
\right)^{-(1+\alpha)}.
\end{eqnarray}
The integral apparently becomes singular for $\alpha\geq 0$. It is
convenient to collect divergent terms in $\Gamma$ functions rather
than in terms of hypergeometric functions. This can be achieved
performing ($1+\beta+[\alpha]$) integrations by parts, if this number
is greater than zero. Defining
\begin{eqnarray}
A(\alpha,\beta) & = & \int\limits_{\frac{t}{t+t'}}^1\! dz\,
z^{\alpha}\, \left( z- \frac{t}{t+t'}\right)^{\beta}\nonumber\\
& = & \frac{1}{1+\beta}\, \frac{t^\alpha\, t'^{\beta+1}}{(t +
t')^{1+\alpha+\beta}} \,\,{}_2F_1(-\alpha,1+\beta,2+\beta,-t'/t)
\end{eqnarray}
we find, if {\bf P} denotes an operator for partial integration:
\begin{eqnarray}
{\bf P}\,A(\alpha,\beta) & = &
\frac{1}{1+\beta}\,(t+t')^{-(1+\beta)}\, t'^{\beta+1} -
\frac{\alpha}{\beta+1}\, A(\alpha-1,\beta+1),\nonumber\\
\frac{d}{dt'}\, A(\alpha,\beta) & = & \beta\, \frac{t}{(t+t')^2}\,
A(\alpha,\beta-1).
\end{eqnarray}
This yields finally (for $(1+\beta+[\alpha])>0$)
\begin{eqnarray}
\lefteqn{\hat{K}(0,0,\alpha\not\in {\bf N},\beta\in {\bf
N},\gamma) =
\frac{(-1)^\beta}{\Gamma(-\alpha)\Gamma(\gamma)}}\nonumber\\
 & \times & t^\alpha t'^\beta\,\frac{d^\beta}{dt'^\beta}\,
\frac{t'^{1+\alpha+\beta-\gamma}}{
(t+t')^{1+\alpha}}\, [{\bf P}]^{1+\beta+[\alpha]}
\int\limits_{\frac{t}{t+t'}}^1\! dz\,
z^{1+\alpha-\gamma}\, \left( z- \frac{t}{t+t'}
\right)^{-(1+\alpha)}.
\end{eqnarray}

\subsection{Second Application of the Borel Transformation}

In terms of the new variables $\tau=1/t$ and $\tau'=1/t'$, a typical
contribution to the Borel transformed amplitude is of the generic
form
\begin{equation}
\frac{1}{\tau^\alpha\tau'^\beta(\tau+\tau')^\gamma}
\end{equation}
with integer $\alpha$, $\beta$, $\gamma$. Some terms are
accompanied by logarithms of $\tau$, $\tau'$ or $\tau+\tau'$, which we
take into account by replacing $\alpha$ by $\alpha + \epsilon$, e.g.,
and expanding around $\epsilon=0$ afterwards. The expressions we are
dealing with have mass dimension five, thus it is feasible by
repeated replacements $\tau\to (\tau+\tau')-\tau'$ to reduce all
powers in the denominator to non--negative values (plus $\epsilon$
for logarithms). We are left with the calculation of (with in general
non--integer values of $\alpha$, $\beta$, $\gamma$)
\begin{equation}
\hat{L}({\alpha},{\beta},{\gamma}) \equiv
\hat{B}_\tau(1/s)\, \hat{B}_{\tau'}(1/s')\,
\frac{1}{\tau^{{\alpha}}\tau'^{{\beta}}
(\tau+\tau')^{{\gamma}}}.
\end{equation}
As usual, we express the denominator in terms of integrals of
exponential functions that have a simple behaviour under
the Borel transformation (cf.\ Eq.~(\ref{eq:Borsimpel})):
\begin{eqnarray}
\lefteqn{\hat{L}({\alpha},{\beta},{\gamma})
=}\nonumber\\
& & =\hat{B}_\tau(1/s)\,\hat{B}_{\tau'}(1/s')\,
\frac{1}{\Gamma({\alpha})\Gamma({\beta})\Gamma({\gamma})}\,
\int\limits_0^\infty \! du_1\, du_2\, du_3\, u_1^{{\alpha}-1}\,
u_2^{{\beta}  -1}\, u_3^{{\gamma}-1}\, e^{-u_1\tau-u_2\tau'
-u_3(\tau +\tau')}\nonumber\\
& & =\frac{s\,s'}{\Gamma({\alpha} )\Gamma({\beta})
\Gamma({\gamma})}\,\left\{\Theta(s'-s)\,
\int\limits_0^s\! du\, u^{{\gamma}-1}\,(s-u)^{{\alpha}
{}-1}\, (s'-u)^{{\beta}-1} + (s \leftrightarrow
s', {\alpha}\leftrightarrow{\beta}) \right\}\makebox[0.6cm]{}
\nonumber\\
& & =\frac{1}{\Gamma({\alpha})\Gamma({\beta})}\, \left\{
s^{{\alpha}  +{\gamma}}\, s'^{ {\beta}}\,
\frac{\Gamma({\alpha})}{\Gamma({\alpha}+{\gamma})}\,\,
{}_2F_1(1-{\beta},{\gamma},{\alpha}+{\gamma},
s/s') + (s \leftrightarrow s', {\alpha}
\leftrightarrow{\beta}) \right\}\label{eq:hyper}.
\end{eqnarray}
As special cases we find
\begin{eqnarray}
\hat{B}_\tau(1/s)\, \hat{B}_{\tau'}(1/s')\,
\frac{1}{\tau^{{\alpha}} (\tau+\tau')^{{\gamma}}} & = &
\frac{ss'}{\Gamma({\alpha})\Gamma({\gamma})}\,
\Theta(s-s')\, (s-s')^{{\alpha}-1}\,
s'^{{\gamma}-1},\\
\hat{B}_\tau(1/s)\, \hat{B}_{\tau'}(1/s')\,\frac{1}{(\tau+
\tau')^{{\gamma}}} & = & \frac{ss'}{\Gamma({\gamma})}\,
s^{{\gamma}-1}\, \delta(s-s').
\end{eqnarray}
Fortunately enough, in the sum of all diagrams most logarithmic terms
cancel and the remaining ones are of type
\begin{equation}
\frac{1}{(\tau+\tau')^{{\gamma}}}\,\ln (\tau+\tau'),
\end{equation}
so we need not expand the hypergeometric function in (\ref{eq:hyper}).

\subsection{A Sample Calculation}\label{app:sample}

In this appendix we calculate the diagrams $D_a$, Fig.~\ref{fig:2}(a),
and $D_b$, Fig.~\ref{fig:2}(b), which contribute to the kinetic
energy. We use
dimensional regularization in $D=4+2\epsilon$ dimensions and work in
Feynman gauge. We use $\Gamma_1 = \Gamma_2 = i\gamma_5$ (cf.\
Eq.~(\ref{CFK})) and do the traces explicitly, which adds a factor two.
Thus, each diagram contributes to the spectral
density $\rho_K$, Eq.~(\ref{SRkin-1}), with a weight factor $1/2$.

In momentum--space, the
lowest order triangle diagram $D_a$ can be expressed as
\begin{equation}
D_a = -6i\int\!\!\frac{d^D\! k}{(2\pi)^D}\,\frac{(vk)
(k^2-(vk)^2)}{k^2(\omega+vk)(\omega'+vk)}\,.
\end{equation}
Since we are interested in the Borel transformed expression only, we
write $vk = (vk+\omega)-\omega$ and keep terms with
non--polynomial dependence on both $\omega$ and $\omega'$:
\begin{equation}
\hat{D}_a = -6i\,\hat{B}_\omega(t)\,\hat{B}_{\omega'}(t')\,
\int\!\!\frac{d^D\! k}{(2\pi)^D}\left(
\omega^2\, \frac{1}{(\omega+vk)(\omega'+vk)} + \omega^3\,
\frac{1}{k^2(\omega+vk)(\omega'+vk)}\right).
\end{equation}
The first loop--integral vanishes, thus we are left with
\begin{equation}\label{eq:A41}
\hat{D}_a = {}-6i\,\hat{B}_\omega(t)\,\hat{B}_{\omega'}(t')
\omega^3\,\int\!\!\frac{d^D\! k}{(2\pi)^D}
\,\frac{1}{k^2(\omega+vk)(\omega'+vk)}.
\end{equation}
The integral encountered here is a special case of the general
one--loop integral (\ref{eq:master1L}):
\begin{equation}
I_1(1,1,1) = \frac{i}{8\pi^{D/2}}\,g_1 g_2
\int\limits_0^1\!\! dx\, (\omega'_E x + \omega_E (1-x))^{D-4},
\end{equation}
where we have changed to the Euclidean variables $\omega_E = -\omega$,
$\omega'_E = -\omega'$. As a shorthand, we use the notations $g_1 =
\Gamma((D-4)/2)$ and
$g_2 = (4-D)\Gamma(4-D)$. Although the integral can
easily be solved in the present case, the above form is convenient
for applying the Borel transformation afterwards. In general, we
have to apply the Borel transformation to terms like
\begin{equation}
\hat{K}(p,q;\alpha,\beta,\gamma) \equiv\,\hat{B}_\omega(t)\,
\hat{B}_{\omega'}(t')\,\omega_E^\alpha\,\omega_E'^\beta
\int\limits_0^1\!\! dx\,
x^p\, (1-x)^q\, (\omega'_E x + \omega_E (1-x))^{-\gamma}.
\end{equation}
With the formul\ae\/ given in App.~\ref{app:A}2 we find
\begin{equation}
\hat{K}(0,0;3,0,4-D) = \frac{1}{g_2}\,(D-4)(D-2)(D-1)D\,t^D\,
t'^D\, (t+t')^{-1-D}.
\end{equation}
Taking altogether, we have
\begin{eqnarray}
\hat{D}_a & = & {}-\frac{3}{4\pi^{D/2}}\,g_1\,(D-4)(D-2)(D-1)D\,
t^D\,t'^D\, (t+t')^{-1-D}\nonumber\\
& = & {}-\frac{36}{\pi^2}\,\frac{1}{tt'}\,\left(
\frac{tt'}{t+t'}\right)^5 + {\cal O}(\epsilon).
\end{eqnarray}
Finally, to calculate the double spectral density of $D_a$, we apply
to $\hat{D}_a$ a Borel transformation once again. More
accurately: if $\rho_a$ is the spectral density, we have
\begin{eqnarray}
\rho_a(s,s') & = & \frac{1}{ss'}\, \hat{B}_\tau(1/s)\,
\hat{B}_{\tau'}(1/s')\,\left[\hat{D}_a\,(s\to 1/\tau,s'\to
1/\tau')\right]\nonumber\\
& = & {}-\frac{3}{4\pi^{D/2}}\,g_1\, \frac{D-4}{\Gamma(D-2)}\,
s^D\,\delta(s-s').
\end{eqnarray}
We have checked explicitly that in $D=4$ dimensions this result agrees
with that obtained by Cutkosky rules.

Next we turn to the two--loop diagram $D_b$ given by
\begin{eqnarray}
\hat{D}_b & = & \hat{B}_\omega(t)\,\hat{B}_{\omega'}(t')\,
32\pi\alpha_s\,\{\,2\omega_E \omega_E'\, (\omega_E+
\omega_E')\,I_2(1,1,1,0,1,1,0)\nonumber\\
& & {} - \frac{1}{2}\,(3\omega_E+\omega_E')\,
I_2(0,1,1,0,1,1,0) -
\frac{1}{2}\,(3\omega_E'+\omega_E)\,I_2(1,0,1,0,1,1,0)
\nonumber\\
& & {}+ \frac{1}{2}\,(\omega_E+\omega_E')\, I_2(1,1,0,0,1,1,0) \}
\end{eqnarray}
with the two--loop integrals
\begin{eqnarray}
\lefteqn{I_2(\alpha,\beta,\gamma,p,q,r,s) =}\nonumber\\
& = & \int\!\! \frac{d^D\! k}{(2\pi)^D}\,\frac{d^D\!
l}{(2\pi)^D}\,\frac{1}{[k^2]^\alpha\,[l^2]^\beta\,[(k-l)^2]^\gamma\,
(\omega+vk)^p\,(\omega+vl)^q\, (\omega'+vk)^r\, (\omega'+vl)^s}\,.
\end{eqnarray}
In the above case all integrals can be related to one--loop integrals,
except for $I_2(1,1,1,0,1,1,0)$. Here it proves useful to
employ the method of integration by parts\cite{BG91,CT81}. From
\begin{equation}
\int\!\! \frac{d^D\! k}{(2\pi)^D}\,\frac{d^D\!
l}{(2\pi)^D}\, \frac{\partial}{\partial l_\mu}\,
\frac{l_\mu}{[k^2]^\alpha
[l^2]^\beta [(k-l)^2]^\gamma (\omega+vl)^q (\omega'+vk)^r} = 0
\end{equation}
we get
\begin{eqnarray}
\lefteqn{(D-2\beta-\gamma-q)\,I_2(\alpha,\beta,\gamma,0,q,r,0)-\gamma\,
I_2(\alpha,\beta-1,\gamma+1,0,q,r,0)}\nonumber\\
& & {} + \gamma\, I_2(\alpha-1,\beta,\gamma+1,0,q,r) + q \omega\,
I_2(\alpha,\beta,\gamma,0,q+1,r,0) =0.
\end{eqnarray}
This relation still contains the term
$I_2(\alpha,\beta,\gamma,0,q+1,r,0)$, which is as difficult to
calculate as the original expression. We can, however, make use of a
corresponding relation that follows from
\begin{equation}
\int\!\! \frac{d^D\! k}{(2\pi)^D}\,\frac{d^D\!
l}{(2\pi)^D}\,\frac{\partial}{\partial l_\mu}\, \frac{k_\mu}{
[k^2]^\alpha
[l^2]^\beta [(k-l)^2]^\gamma (\omega+vl)^q (\omega'+vk)^r} = 0
\end{equation}
to eliminate $I_2(\alpha,\beta,\gamma,0,q+1,r,0)$ and get
\begin{eqnarray}
\lefteqn{I_2(\alpha,\beta,\gamma,0,q,r,0) =
\frac{1}{\beta-\gamma-\frac{\omega'}{\omega}\,(2\beta+\gamma+q-D)}\,
\{\beta\,I_2(\alpha,\beta+1,\gamma-1,0,q,r,0)} \nonumber\\
& & {}-\beta\, I_2(\alpha-1,\beta+1,\gamma,0,q,r,0) -
q I_2(\alpha,\beta,\gamma,0,q+1,r-1,0)\nonumber\\
& & {}+ \gamma \left(1-\frac{\omega'}{\omega}\right)\,
{}[\, I_2(\alpha-1,\beta,\gamma+1,0,q,r,0) -
I_2(\alpha,\beta-1,\gamma+1,0,q,r,0)\, ]\, \}.
\end{eqnarray}
Using this expression, one can relate $\hat{D}_b$ to $\hat{K}$,
and apply the second Borel transformation according to the
formul\ae\/ given in App.~\ref{app:A}3. Note that both the diagram
and its double spectral
function are finite in the limit $D\to 4$, whereas all the other
two--loop diagrams in Fig.~\ref{fig:2} have to be renormalized. The
calculation of the remaining diagrams proceeds along the same lines.
The results are given in App.~\ref{app:B}.

\section{Expressions for the two--loop diagrams}\label{app:B}

\begin{figure}
$$
\ \psfig{file=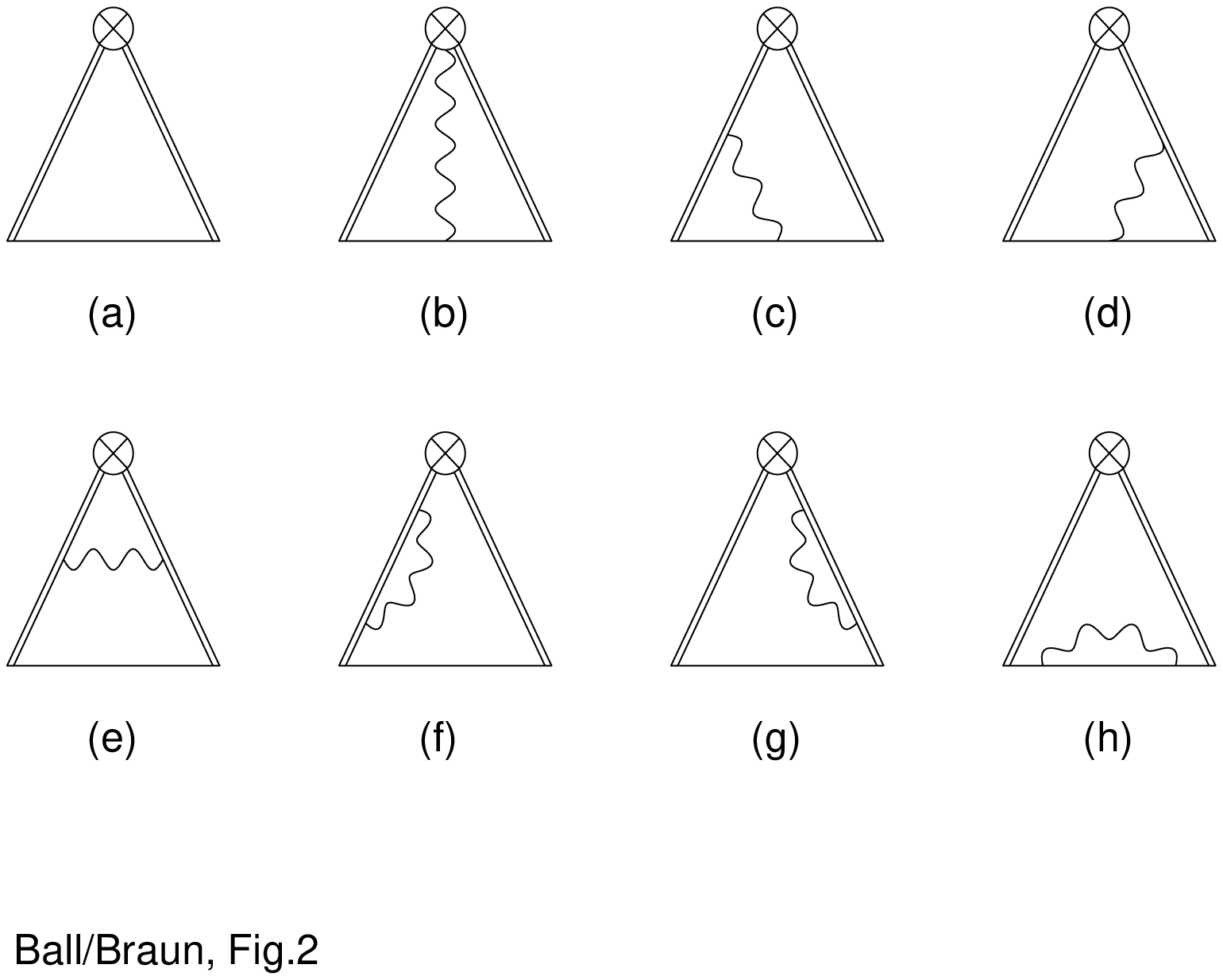,height=7in}
$$
\caption{The diagrams contributing to $\rho_K^{\text{pert}}$,
Eq.~(\protect{\ref{pert-2loop}}).\label{fig:2}}
\end{figure}

\begin{figure}
$$
\ \psfig{file=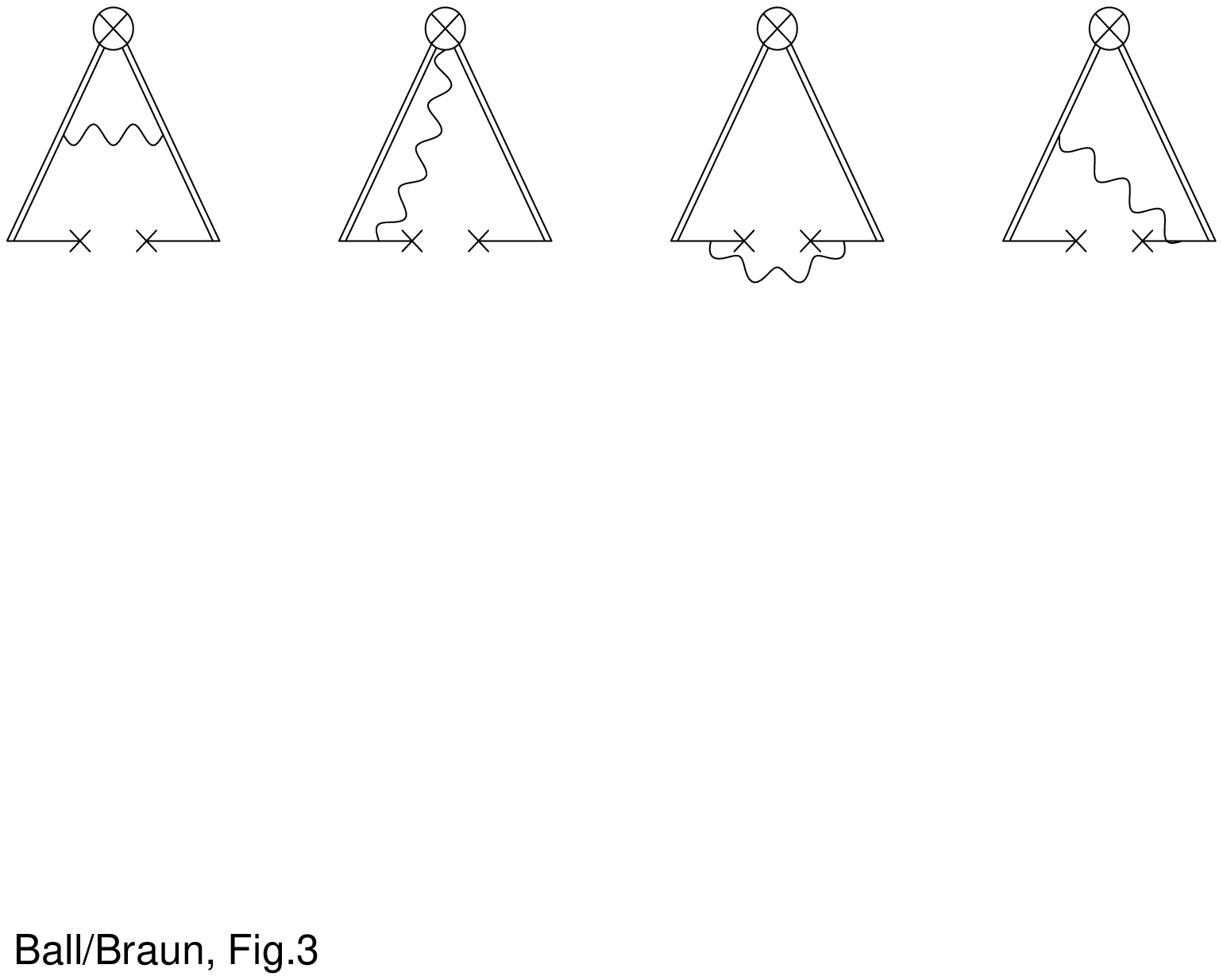,height=7in}
$$
\caption{The diagrams contributing to $\rho_K^{(3)}$,
Eq.~(\protect{\ref{K3}}). \label{fig:3}}
\end{figure}

\begin{figure}
$$
\ \psfig{file=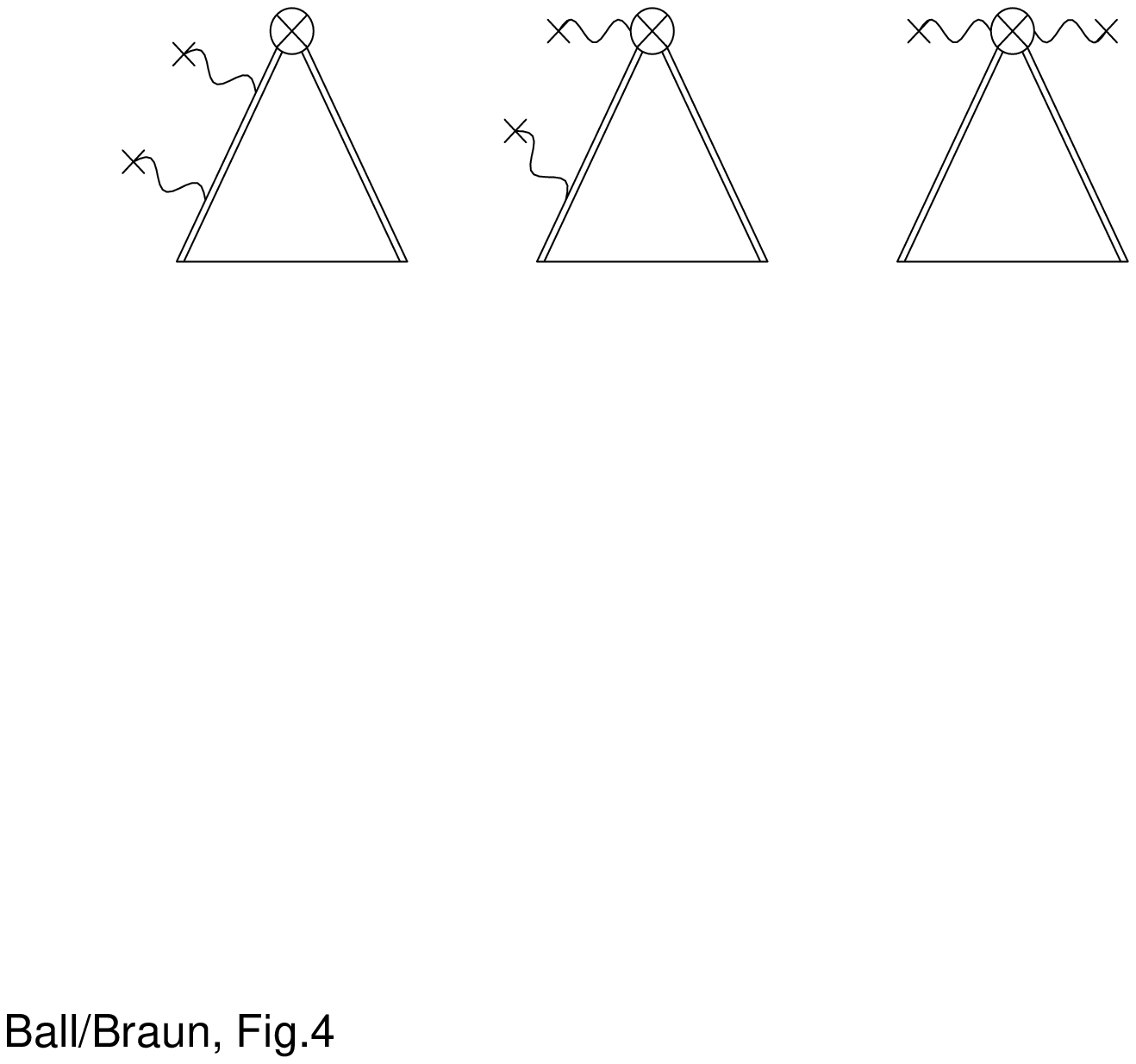,height=7in}
$$
\caption{The diagrams contributing to $\rho_K^{(4)}$,
Eq.~(\protect{\ref{K4}}), in Fock--Schwinger gauge. The lower right
vertex is put to the origin in coordinate space.\label{fig:4}}
\end{figure}

\begin{figure}
$$
\ \psfig{file=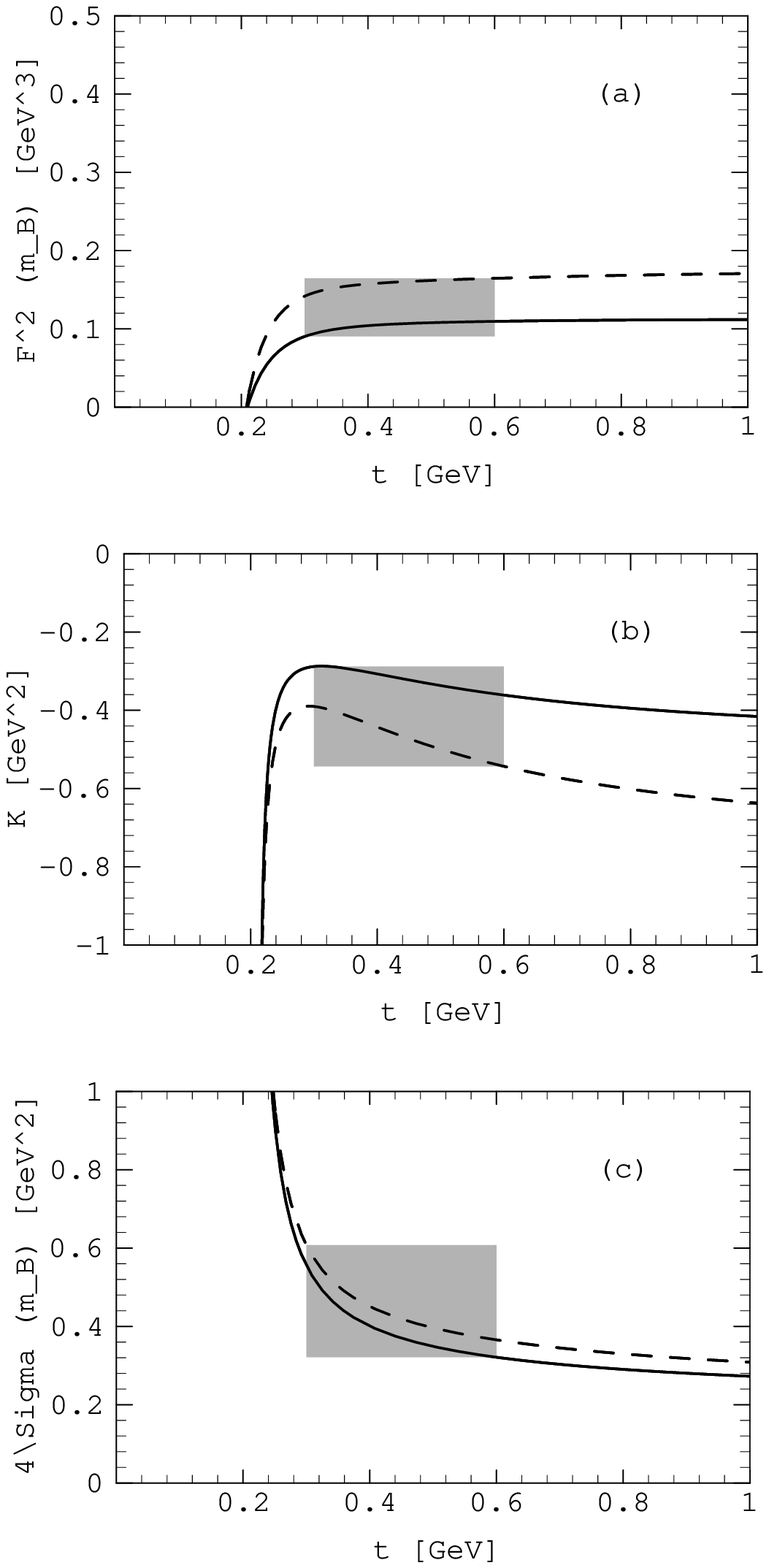,height=7.5in}
$$
\caption{The sum rules (\protect{\ref{SR}}) to leading--order
accuracy (i.e.\ using the spectral densities (\protect{\ref{spectr}}))
as functions of the Borel parameter $t$ for different values of the
continuum threshold $\omega_0$ (solid lines: $\omega_0 =
1.0\,\text{GeV}$ [$\bar\Lambda=0.4\,\text{GeV}$], dashed lines:
$\omega_0 = 1.2\,\text{GeV}$ [$\bar\Lambda=0.5\,\text{GeV}$]). The
shaded areas indicate the working regions of the sum rules.
(a) $F^2(\mu = m_B)$, (b) the kinetic energy $K$, (c) $m_V^2-m_P^2 =
4\Sigma$ at the scale $\mu = m_B$.\label{fig:5}}
\end{figure}

\begin{figure}
$$
\ \psfig{file=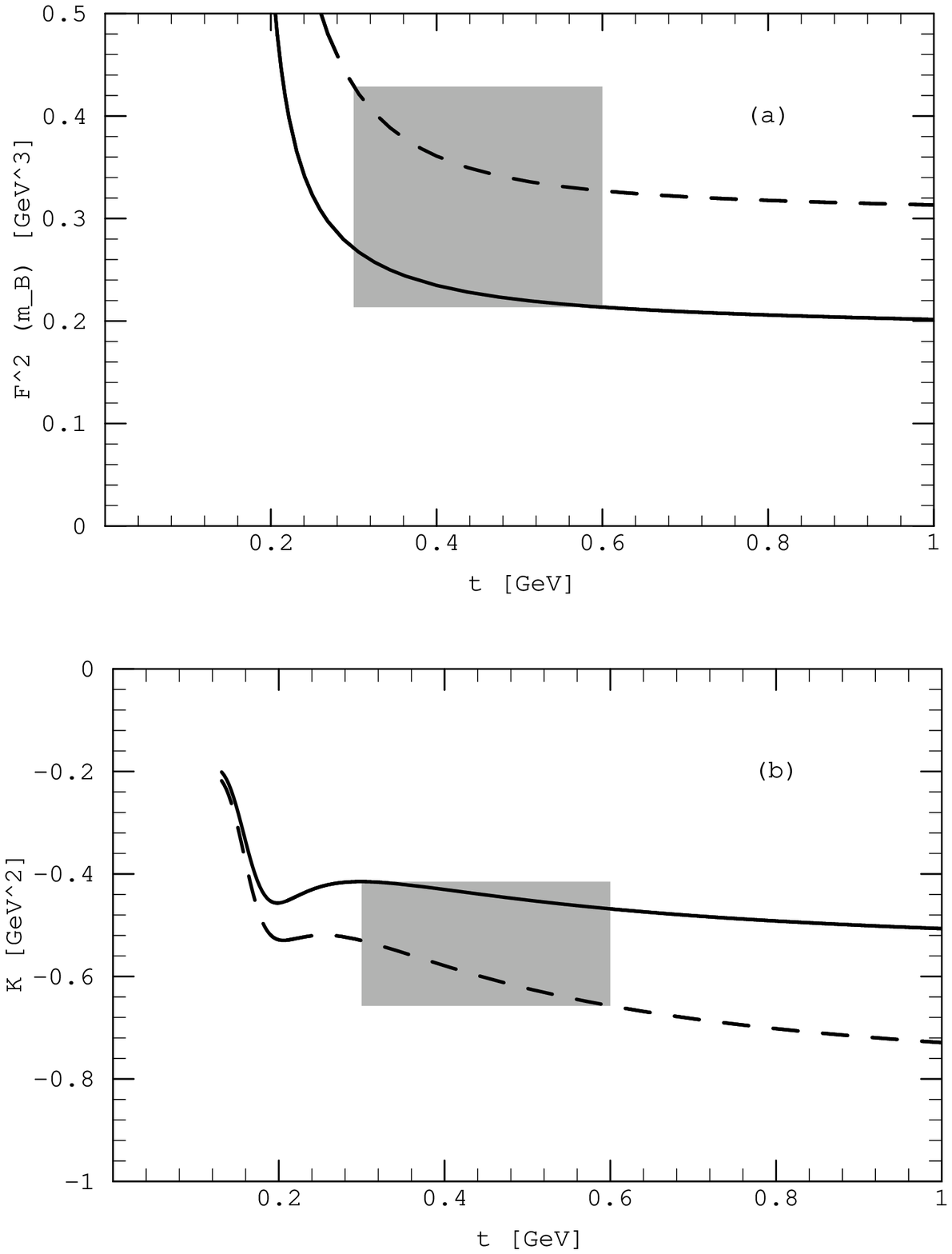,height=7.5in}
$$
\caption{The sum rules (\protect{\ref{SRFB}}) and
(\protect{\ref{eq:rhoK2L}}) in next--to--leading order
accuracy. The parameter values are the same as in
Fig.~\protect{\ref{fig:5}}. (a)
$F^2(\mu = m_B)$, (b) the kinetic energy $K$.\label{fig:6}}
\end{figure}

In this appendix we collect the explicit expressions for the
renormalized Feynman diagrams in Fig.~\ref{fig:2}, after
application of the Borel transformation in both external momenta.
The missing diagrams with a gluon line connecting the operator
vertex with a
heavy quark line vanish in the limit of equal heavy quark velocities.
$\hat{D}_a$, $\hat{D}_b$, \dots, denote the Borel transformed
diagrams in Fig.~\ref{fig:2}(a), (b), \dots, respectively. Each diagram
contributes with a weight $1/2$ to the spectral density
Eq.~(\ref{pert-2loop}).
\begin{eqnarray}
\hat{D}_a^{\text{pert}} & = & -\frac{36}{\pi^2}\,\frac{1}{tt'}\,
\left(\frac{tt'}{t+t'} \right)^5,\\
\hat{D}_b^{\text{pert}} & = & \frac{\alpha_s(\mu)}{\pi}\,
\frac{2tt'}{(t+t')^4 \pi^2}\, \left\{ tt' (t+t')^3 - 2 t^3 \ln
\frac{t}{t+t'}\, (t^2 + 5 tt' + 10t'^2) \right.\nonumber\\
& & \phantom{\frac{\alpha_s(\mu)}{\pi}\,
\frac{2tt'}{(t+t')^4 \pi^2}\: } \left. {}- 2 t'^3 \ln \frac{t'}{t+t'}
\,(t'^2 + 5 tt' + 10t^2) \right\},\\
\hat{D}_{c+d}^{\text{pert}} & = & \frac{\alpha_s(\mu)}{\pi}\,
\frac{2tt'}{(t+t')^5 \pi^2}\,\left\{ -tt'(t+t')^4 - 12 t^2 t'^2
(t+t')^2\right.\nonumber\\
& & {} + 24 t^3 t'^3 \left(\frac{3}{4} - \frac{\pi^2}{3} + \ln 2 +
 \ln \frac{t+t'}{\mu} \right)  + 2 t^3 \ln \frac{t}{t+t'} \,
(t^3 + 6t^2 t' + 15 t t'^2 + 22 t'^3)\nonumber\\
& & \left. {} + 2 t'^3 \ln \frac{t}{t+t'} \, (t'^3 + 6t'^2 t + 15 t'
t^2 + 22 t^3) \right\},\\
\hat{D}_e^{\text{pert}} & = & \frac{\alpha_s(\mu)}{\pi}\,
\frac{4}{\pi^2}\,\frac{t^2t'^2}{(t+t')^2}\, \left\{ -(t+t')^4 + tt'
(t+t')^2\right.\nonumber\\
& & \left.\phantom{\frac{\alpha_s(\mu)}{\pi}\,\frac{4}{\pi^2}\,
\frac{t^2t'^2}{(t+t')^2}\:} + 12 t^2
t'^2 \left(\frac{5}{4} + \gamma_E - \ln 2 - \ln \frac{t+t'}{\mu}
\right) \right\},\\
\hat{D}_f^{\text{pert}} & = & \frac{\alpha_s(\mu)}{\pi}\,
\frac{48}{\pi^2}\, \frac{t^4 t'^4}{(t+t')^5}\, \left(-1-\gamma_E +
\ln 2 + \ln \frac{t}{\mu} \right),\\
\hat{D}_g^{\text{pert}} & = & \hat{D}_f^{\text{pert}}
(t\leftrightarrow t'),\\
\hat{D}_h^{\text{pert}} & = & \frac{\alpha_s(\mu)}{\pi}\,
\frac{6}{\pi^2}\, \frac{t^4 t'^4}{(t+t')^5}\, \left( -1 + 4\gamma_E -
4 \ln 2 - 4 \ln \frac{tt'}{\mu(t+t')}\right).
\end{eqnarray}

\newpage
\newpage
\newpage
\newpage
\newpage
\end{document}